\documentclass[aps,prc,twocolumn,nofootinbib,showpacs,floatfix]{revtex4}
\usepackage{epsfig}
\usepackage{bm}
\usepackage{dcolumn}

\usepackage[normalem]{ulem}  
\usepackage[dvips]{color} 

\def\bea {\begin{eqnarray}}
\def\eea {\end{eqnarray}}
\def\be {\begin{equation}}
\def\ee {\end{equation}}
\def\ben{\begin{enumerate}}
\def\een{\end{enumerate}}
\def\bi{\begin{itemize}}
\def\ei{\end{itemize}}

\def\etal{{\it et al.}\ }

\def\O{{\cal O}}
\def\F{{\cal F}}
\def\A{{\cal A}}
\def\prl {Phys. Rev. Lett.\ }
\def\pl {Phys. Lett.\ }
\def\pr {Phys. Rev.\ }
\def\np {Nucl. Phys.\ }

\def\gA{g_{\mbox{\tiny A}}}

\def\GV{G_{\mbox{\tiny V}}}
\def\GF{G_{\mbox{\tiny F}}}
\def\DRV{\Delta_{\mbox{\tiny R}}^{\mbox{\tiny V}}}
\def\mW{m_{\mbox{\tiny W}}}
\def\mA{m_{\mbox{\tiny A}}}
\def\mZ{m_{\mbox{\tiny Z}}}
\def\hyphen{{\mbox{-}}}

\newcommand{\sfrac}[2]{\mbox{\small{$\frac{#1}{#2}$}}}

\def\2p{|2p\rangle }
\def\4p2h{|4p\hyphen 2h\rangle }
\def\6p4h{|6p\hyphen 4h\rangle }
\def\2h{|2h\rangle }
\def\4h2p{|4h\hyphen 2p\rangle }
\def\6h4p{|6h\hyphen 4p\rangle }

\begin{document} 
\preprint{ }
 
\title{An improved calculation of the isospin-symmetry-breaking corrections
to Superallowed Fermi $\beta$ decay}

\author{I.S. Towner}
\altaffiliation{Present address: 
Department of Physics,
Queen's University, Kingston, Ontario K7L 3N6, Canada}
\author{J.C. Hardy}
\affiliation{Cyclotron Institute, Texas A \& M University,                    
College Station, Texas  77843}
\date{\today} 
\begin{abstract} 
We report new shell-model calculations of the isospin-symmetry-breaking correction,
$\delta_C$, to superallowed $0^{+} \rightarrow 0^{+}$ nuclear $\beta$ decay.  The
most important improvement is the inclusion of core orbitals, which are demonstrated
to have a significant impact on the mismatch in the radial wave functions of the
parent and daughter states.  We determine which core orbitals are important to
include from an examination of measured spectroscopic factors in single-nucleon
pick-up reactions.  In addition, where new sets of effective interactions have
become available since our last calculation, we now include them; this leads
to small changes in $\delta_{NS}$ as well.  We also examine the new radiative-correction
calculation by Marciano and Sirlin and, by a simple reorganization, show that it is
possible to preserve the conventional separation into a nucleus-independent ``inner"
radiative term, $\DRV$, and a nucleus-dependent ``outer" term, $\delta_R^{\prime}$
We tabulate the new values for $\delta_C$, $\delta_{NS}$ and $\delta_R^{\prime}$ for
twenty superallowed transitions, including the thirteen currently well-studied cases.
With these new correction terms the corrected $\F t$ values for the thirteen cases
are statistically consistent with one another and the anomalousness of the $^{46}$V
result disappears.  These new calculations lead to a lower average $\overline{\F t}$
value and a higher value for $V_{ud}$.  The sum of squares of the top-row elements
of the CKM matrix now agrees exactly with unitarity.           

\end{abstract} 


\maketitle

\section{Introduction}
\label{s:intro}

Superallowed $0^{+} \rightarrow 0^{+}$ nuclear $\beta$ decay currently
provides the most precise value for $V_{ud}$, the up-down element
of the Cabibbo-Kobayashi-Maskawa (CKM) matrix \cite{HT05,HT05a,Ha07}. 
This element is the key ingredient of the most demanding available test
of CKM-matrix unitarity, a fundamental requirement of the electroweak
standard model.  To extract $V_{ud}$ from the experimental data, small
theoretical corrections -- of order $\sim$1\% -- must be applied to take
account of unobserved radiative effects as well as the isospin symmetry-breaking
that occurs between the analog parent and daughter states of each superallowed
transition \cite{TH02,OB95}.  Even though these corrections are very small, 
experimental measurements have by now reached such high precision that the
uncertainty on $V_{ud}$ ($\pm$0.03\%) is currently dominated not by experiment
but by the uncertainty on these theoretical corrections.

In the determination of $V_{ud}$, an important strength of the nuclear
measurements is that there are many $0^{+} \rightarrow 0^{+}$ transitions
available for study, and currently there are thirteen of them, ranging
from $^{10}$C to $^{74}$Rb, that have been measured with high precision.
With so many, it becomes possible to validate the analysis procedure by
checking that all transitions individually yield statistically consistent
results for $V_{ud}$.  Since the isospin-symmetry-breaking corrections
depend on nuclear structure, they differ from transition to transition
and are particularly sensitive to this consistency test.  Thus the
appearance of an anomalous result from any transition could signal a
problem with the structure-dependent correction for that case, a problem
which might have implications for other cases as well.

In the most recent survey of superallowed $0^{+} \rightarrow 0^{+}$
transitions, which appeared in 2005 \cite{HT05}, the results for all
precisely measured cases -- there were twelve at that time -- were
statistically consistent with one another.  Today, there are thirteen
such cases and they still form a statistically consistent ensemble
overall.  However, recent precise Penning-trap measurements \cite{Sa05,Er06b}
of the $Q_{EC}$ value for the superallowed decay of $^{46}$V have left
the result for that transition more than two standard deviations away
from the average of all other well-known transitions.  This possible anomaly
led us initially to reexamine the isospin-symmetry-breaking corrections for the
$^{46}$V transition, but what we learned from that reexamination prompted
us to a more general reevaluation of the corrections for other transitions
as well.

Our previous shell-model calculations for $^{46}$V considered six valence
nucleons occupying the $pf$-shell orbitals outside a $^{40}$Ca closed
shell.  This model space generated reasonable energies and spins for the
known states in $^{46}$Ti, the daughter of $^{46}$V.  However, an important
part of the charge-dependent correction depends on the radial mismatch
between the decaying proton in the parent nucleus and the resulting neutron
in the daughter nucleus; but both these nucleons are bound to $^{45}$Ti, so
the structure of that nucleus turns out to be important too.  What is most
striking about $^{45}$Ti is that it has a $3/2^+$ state at an excitation energy
of only 330 keV, which is strongly populated in single-nucleon pick-up reactions
like $(p,d)$ and ($^3$He,$\alpha$).  Such low-lying $sd$-shell states can contribute
to the structural parentage of the initial and final states of the superallowed
transition and consequently must affect the radial mismatch between them.
This indicated to us that a complete calculation of the isospin-symmetry-breaking
correction for the decay of $^{46}$V should include contributions from shells
deeper than the $pf$ shell. 
   
Two questions then arose.  How many deeper shells need to be included and,
if this effect is important for $^{46}$V decay, how many other transitions
will be similarly affected?  In section \ref{s:isbc} of this paper, we address
these questions and settle on criteria for including deeper shells.  Using these
criteria -- and incorporating more recent effective interactions that have
become available since our last work -- we then re-evaluate the
isospin-symmetry-breaking corrections for all
transitions of relevance to the study of superallowed $0^{+} \rightarrow 0^{+}
\beta$ decay.  For the cases with $A \leq 38$ the changes in the corrections
are very small -- typically 0.03\% -- but for the heavier nuclei the changes
can be as large as 0.2\%.  Most significantly, with the new calculated corrections,
the result for $^{46}$V is no longer anomalous. 

In section \ref{s:radc}, we incorporate recent improvements made by Marciano and
Sirlin \cite{MS06} to the calculation of the radiative corrections for superallowed
decays and then in section \ref{s:Ft} we apply both types of corrections --
isospin-symmetry-breaking and radiative -- to the current experimental data for
superallowed decays.  The result for $V_{ud}$ is changed appreciably, although it
is still within quoted uncertainties of its old value, and the CKM-unitarity sum
is improved.

\section{Superallowed beta decay}
\label{s:sbd}

Superallowed Fermi beta decay between $0^+$ states depends uniquely
on the vector part of the hadronic weak interaction.  When it occurs
between isospin $T=1$ analog states, the conserved vector current (CVC)
hypothesis indicates that the $ft$ values should be the same
irrespective of the nucleus, {\it viz.}
\be
ft = \frac{K}{\GV^2 | M_F |^2} = {\rm ~const},
\label{ftconst}
\ee
where $K/(\hbar c )^6 = 2 \pi^3 \hbar \ln 2 / (m_e c^2)^5 =
( 8120.278 \pm 0.004 ) \times 10^{-10}$ GeV$^{-4}$s; $\GV $ is
the vector coupling constant for semi-leptonic weak interactions;
and $M_F$ is the Fermi matrix element.  The CVC hypothesis
asserts that the vector coupling constant, $\GV$, is a true constant
and not renormalised to another value in the nuclear medium.

In practice, Eq.\,(\ref{ftconst}) has to be amended slightly.  Firstly, 
there are radiative corrections because, for example, the emitted
electron may emit a bremsstrahlung photon that goes undetected
in the experiment.  Secondly, isospin is not an exact symmetry in nuclei
so the nuclear matrix element, $M_F$, is slightly reduced from its ideal
value, leading us to write:
\be
|M_F|^2 = |M_0|^2 ( 1 - \delta_C ) ,
\label{MF2}
\ee
where $M_0$ is the exact-symmetry value, which for $T = 1$ states
is $M_0 = \sqrt{2}$.  Thus, we define a ``corrected" $\F t$ value as
\be
\F t \equiv ft (1 + \delta_R )(1 - \delta_C ) = \frac{K}{2 \GV^2 
(1 + \DRV )} = {\rm ~const},
\label{Ftconst}
\ee
where $\delta_C$ is the isospin-symmetry-breaking correction,
$\delta_R$ is the transition-dependent part of the radiative
correction, and $\DRV$ is the transition-independent part. 
Fortunately these corrections are all of order 1\% but, even so, to
maintain an accuracy criterion of 0.1\% they must be calculated
with an accuracy of 10\% of their central value.  This is a demanding
request, especially for the nuclear-structure-dependent corrections.

To separate out those terms that are dependent on nuclear
structure from those that are not, we split the transition-dependent radiative
correction into two terms,
\be
\delta_R = \delta_R^{\prime} + \delta_{NS},
\label{dr}
\ee
of which the first, $\delta_R^{\prime}$, is a function only of the electron's
energy and the charge of the daughter nucleus $Z$; it therefore
depends on the particular nuclear decay, but is {\em independent} of
nuclear structure.  The second term, $\delta_{NS}$, like $\delta_C$, 
depends in its evaluation on the details of nuclear structure.
To emphasize the different sensitivities of the correction terms,
we rewrite the expression for $\F t$ as
\be
\F t \equiv ft (1 + \delta_R^{\prime}) (1 + \delta_{NS} - \delta_C ) =
\frac{K}{2 \GV^2(1 + \DRV )},
\label{Ftfactor}
\ee
where the first correction in brackets is independent of nuclear
structure, while the second incorporates the structure-dependent terms.

From Eq.\,(\ref{Ftfactor}) it can be seen that a measurement of any
one superallowed transition establishes a single value for $\GV$;
moreover, measurements of many transitions provides an excellent test
of the validity of the whole analysis.  Since CVC requires a unique
value of $\GV$, all the extracted $\F t$-values should be identical
within experimental uncertainties.

\begin{figure*}[t]
\epsfig{file=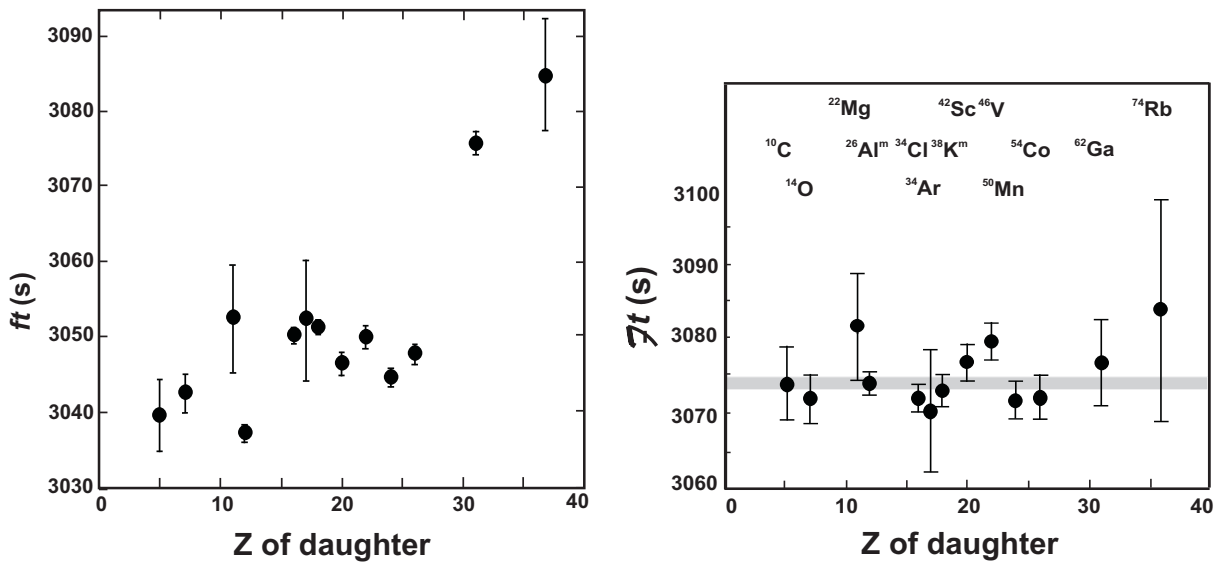,width=16cm}
\caption{Results from the 2005 survey \cite{HT05} updated with more recent
published results \cite{Sa05, To05, Hy05, Er06a, Er06b, Bo06, Ba06, Ia06, Hy06, Bu06}.
The uncorrected $ft$ values for the thirteen best known superallowed decays (left)
are compared with the same results after corrections have been applied to
obtain $\F t$ values.  Here we have used the corrections calculated by us in
2002 \cite{TH02}, which were used in the original survey.   The shaded horizontal
band gives one standard deviation around the average $\F t$ value.}
\label{fig1}
\end{figure*}

The $ft$-value that characterizes any $\beta$-transition depends on
three measured quantities: the total transition energy, $Q_{EC}$; the
half-life, $t_{1/2}$, of the parent state; and the branching ratio,
$R$, for the particular transition of interest.  The $Q_{EC}$-value
is required to determine the statistical rate function, $f$, while the
half-life and branching ratio combine to yield the partial half-life, $t$.
In 2005 we published a new survey of world data on superallowed
$0^{+}\!\rightarrow\!0^{+}$ beta decays \cite{HT05}.  All previously
published measurements were included, even those that were based on
outdated calibrations if enough information was provided that they
could be corrected to modern standards.  In all, more than 125
independent measurements of comparable precision, spanning four
decades, made the cut.  In the two years since the survey was closed
another ten relevant publications have appeared \cite{Sa05, To05, Hy05, Er06a, 
Er06b, Bo06, Ba06, Ia06, Hy06, Bu06} and we have now incorporated these
results into our data base.  Based on these data for the thirteen most
precisely known transitions, we obtain the $ft$ values shown on the left
side of Figure~\ref{fig1}; then, by incorporating the corrections calculated
by us in 2002 \cite{TH02} and used in our 2005 survey \cite{HT05}, we
obtain the corrected $\F t$ values plotted on the right side of the figure.

Obviously the calculated corrections do a remarkable job eliminating the
considerable scatter that is evident in the $ft$-value plot on the left but
is absent in the corrected $\F t$ values shown on the right.  Overall, the
statistical agreement among the $\F t$ values is quite satisfactory, the
normalized $\chi^2$ being 0.8.  Thus, considering that the correction terms
were evaluated completely independently of these data, the consistency among
the $\F t$ values can be taken as strong evidence that the correction terms
are, in general, soundly based.

However, there is a small but noticeable
deviation from the average at $^{46}$V (and possibly $^{42}$Sc), which has
only been revealed by the recent Penning-trap measurements \cite{Sa05,Er06b}
of the transition $Q_{EC}$ values.  Though its statistical significance
appears rather marginal in the figure, it must be remarked that the uncertainties quoted
on these $\F t$ values have been very conservatively determined.  The measured
data for each input parameter -- $Q_{EC}$-value, half-life and branching ratio -- 
were separately evaluated \cite{HT05} and, if the measurements were inconsistent
with one another, the weighted-average uncertainty for that parameter was
increased to account for that inconsistency.  In effect, for such cases, the
original uncertainties quoted with the published measurements were all increased
by a common ``scale factor" that was large enough to restore statistical consistency
among the measurements.  (These scale factors are tabulated for each parameter in
Ref.\,\cite{HT05}; they range from 1 to 3.6.)  This method, which is also used by
the Particle Data Group \cite{PDG}, leads to final average values that have a high
confidence level but it does so at the cost of producing uncertainties that are
in many cases larger than would result from a strict statistical average.

With this method of analysis in mind, the excursion of the $^{46}$V $\F t$ value
cannot be entirely ignored as a possible signal that the nuclear-structure-dependent
corrections in this mass region are deficient.  It certainly proved to be sufficiently
provocative that we were led to the reevaluation of correction terms that is
reported here.

\section{Isospin-symmetry breaking correction, $\delta_C$}
\label{s:isbc}

For weak vector interactions in hadron states, the CVC hypothesis
protects the decay amplitudes from
strong-interaction corrections.  However, there is a caveat.
The CVC hypothesis also requires the hadron state to be an exact
eigenstate of $SU(2)$ symmetry (isospin).  In nuclei, $SU(2)$ 
is always broken, albeit weakly, by Coulomb interactions between
protons.  There may be other charge-dependent effects as well.
These influences shift the value of the hadron matrix element
from its exact symmetry limit to a new value and this shift
has to be evaluated before weak-interaction physics can be
probed with hadrons.  In the case of superallowed $\beta$
decay, the hadron matrix element, $M_F$, is given by Eq.\,(\ref{MF2})
and it is $\delta_C$ that we seek to evaluate.

In the shell model for the cases of interest here, the
$A$-particle wave functions representing the initial and final
states for superallowed $\beta$ decay,
$| i \rangle$ and
$| f \rangle$, 
are states of angular momentum zero and isospin one.
In a second quantisation formulation, the Fermi matrix element
is written
\be
M_F = \langle f | \tau_+ | i \rangle = \sum_{\alpha , \beta }
\langle f | a_{\alpha}^{\dag} a_{\beta} | i \rangle
\langle \alpha | \tau_+ | \beta \rangle ,
\label{MF2q}
\ee
where the operator for Fermi $\beta$ decay is the isospin ladder
operator, $a_{\alpha}^{\dag}$ creates a neutron in quantum
state $\alpha$ and $a_{\beta}$ annihilates a proton in
quantum state $\beta$.  The single-particle matrix element,
$\langle \alpha | \tau_+ | \beta \rangle$,
is just a radial integral
\be
\langle \alpha | \tau_+ | \beta \rangle
= \delta_{\alpha , \beta} \int_0^{\infty}
R_{\alpha}^n(r)
R_{\beta}^p(r) r^2 ~dr
\equiv  \delta_{\alpha , \beta} ~ r_{\alpha}.
\label{radi}
\ee
If the proton and neutron radial functions
$R_{\alpha}^n(r)$ and
$R_{\beta}^p(r)$ are identical, then the radial integral reduces
to the normalization integral and has the value $r_{\alpha} =1$.

Now we introduce into Eq.\,(\ref{MF2q}) a complete set of states
for the $(A-1)$-particle system, $|\pi \rangle $, by writing
\be
M_F = \sum_{\pi , \alpha}
\langle f | a_{\alpha}^{\dag} | \pi \rangle
\langle \pi | a_{\alpha} | i \rangle
r_{\alpha}^{\pi} .
\label{MFpar}
\ee
This is the essence of our model: we have allowed the radial
integral to depend on the parentage expansion.
Thus, we have added an additional label
to $r_{\alpha}$ and now write $r_{\alpha}^{\pi}$.

If isospin is an exact symmetry, then the matrix elements of the
creation and annihilation operators are related by hermiticity,
$\langle \pi | a_{\alpha} | i \rangle =
\langle f | a_{\alpha}^{\dag} | \pi \rangle^{\ast} $.
With that requirement, and with the radial integrals set to unity, 
the symmetry-limit matrix element is
\be
M_0 = \sum_{\pi , \alpha} |
\langle f | a_{\alpha}^{\dag} | \pi \rangle |^2 .
\label{M0}
\ee
Thus we see that the breakdown of isospin symmetry can enter the
evaluation of $M_F$ in one of two ways:  either the matrix elements
of $a_{\alpha}$ and $a_{\alpha}^{\dag}$ are not related by
hermiticity, or the radial integrals are not unity.  Since each
effect is small, we can, to first order, write the isospin-symmetry
breaking correction as the sum of two terms
\be
\delta_C = \delta_{C1} + \delta_{C2}
\label{dc1and2}
\ee
where in evaluating $\delta_{C1}$ all radial integrals are set to
unity but the matrix elements are not assumed to be related by hermiticity,
while in evaluating $\delta_{C2}$ it is assumed that
$\langle \pi | a_{\alpha} | i \rangle =
\langle f | a_{\alpha}^{\dag} | \pi \rangle^{\ast} $
but the radial integrals are allowed to differ from unity.  Past 
calculations \cite{TH02,OB95} have indicated the radial overlap
correction, $\delta_{C2}$, is the larger of the two corrections
so we will study this first.

\subsection{Radial Overlap Correction, $\delta_{C2}$}
\label{ss:roc}

\subsubsection{Strategy for calculation}
\label{sss:st}

For the $\delta_{C2}$ calculation, the Fermi matrix element is
\bea
M_F & = & \sum_{\pi , \alpha} |
\langle f | a_{\alpha}^{\dag} | \pi \rangle |^2 r_{\alpha}^{\pi}
\nonumber \\
 & = & \sum_{\pi , \alpha} 
|\langle f | a_{\alpha}^{\dag} | \pi \rangle |^2 
- \sum_{\pi , \alpha} 
|\langle f | a_{\alpha}^{\dag} | \pi \rangle |^2 (1 - r_{\alpha}^{\pi} )
\nonumber \\
& = & M_0 \left ( 1 - \frac{1}{M_0}
\sum_{\pi , \alpha} 
|\langle f | a_{\alpha}^{\dag} | \pi \rangle |^2 \Omega_{\alpha}^{\pi} 
\right )
\label{MF1}
\eea
where $M_0$ is the exact-symmetry value, Eq.\,(\ref{M0}), and
$\Omega_{\alpha}^{\pi}$ has been introduced as a radial-mismatch factor
\be
\Omega_{\alpha}^{\pi} = 
(1 - r_{\alpha}^{\pi} ) .
\label{rmf}
\ee
Recalling that $\delta_{C2}$ is defined as
$|M_F|^2 = |M_0|^2 ( 1 - \delta_{C2} ) $  we obtain
\be
\delta_{C2} \simeq \frac{2}{M_0}
\sum_{\pi , \alpha} 
|\langle f | a_{\alpha}^{\dag} | \pi \rangle |^2 \Omega_{\alpha}^{\pi} 
\label{dc2_1}
\ee
to first order in small quantities.  A large contribution to $\delta_{C2}$
therefore requires a large spectroscopic amplitude and a significant
departure of the radial integral from unity.

There is an opportunity here to take guidance from experiment.  The
square of each spectroscopic amplitude,
$ |\langle f | a_{\alpha}^{\dag} | \pi \rangle |^2 $, 
is related to the spectroscopic factor measured in neutron pick-up
direct reactions.  The exact relation, after inserting the isospin
angular momentum couplings, is
\be
\delta_{C2} \simeq \sum_{\pi, \alpha}
\frac{T_f(T_f+1)+ \sfrac{3}{4} -T_{\pi}(T_{\pi}+1)}{T_f(T_f+1)}
~ S_{\alpha , T_f}^{T_{\pi}} ~ \Omega_{\alpha}^{\pi}
\label{dc2_2}
\ee
where $S_{\alpha , T_f}^{T_{\pi}}$ is the spectroscopic factor for
pick up of a neutron in quantum state $\alpha$ from an $A$-particle
state of isospin $T_f$ to an $(A-1)$-particle state of isospin $T_{\pi}$.
On setting $T_f = 1$ and separately identfying sums to the
isospin-lesser states with $T_{\pi} = \sfrac{1}{2}$, denoted $\pi^<$,
and the isospin-greater states with
$T_{\pi} = \sfrac{3}{2}$, denoted $\pi^>$,  we obtain a very
revealing formula
\be
\delta_{C2} \simeq \sum_{\pi^< , \alpha } S_{\alpha}^< ~ \Omega_{\alpha}^<
- \frac{1}{2} \sum_{\pi^> , \alpha } S_{\alpha}^> ~ \Omega_{\alpha}^> .
\label{dc2_3}
\ee

\begin{table*}[ht]
\begin{center}
\caption{Illustration of the strategy used in calculating $\delta_{C2}$ for $^{46}$V.  The measured spectroscopic
factors from the $^{46}$Ti$(^3$He$,\alpha )^{45}$Ti reaction \cite{Bo67} are shown for the states where they are largest.  Two
calculations are then given for each state's contribution to $\delta_{C2}$: the first assumes that the total
Macfarlane-French (M-F) sum rule is exhausted in each state, while the second gives the result of a complete
shell-model calculation.  Both methods give remarkably similar results.
\label{t:Sfactor}}
\vskip 1mm
\begin{ruledtabular}
\begin{tabular}{ccccccccc}
& & & & & & & & \\[-2mm]
& & &
\multicolumn{1}{c}{$(^3$He$,\alpha )$} & &
\multicolumn{2}{c}{Limiting case} &
\multicolumn{2}{c}{Shell Model} \\
\cline{6-7}
\cline{8-9} \\[-3mm]
\multicolumn{1}{c}{$^{45}$Ti} & & &
\multicolumn{1}{c}{measured \cite{Bo67}} & &
\multicolumn{1}{c}{M-F} &
\multicolumn{1}{c}{contribution} & & 
\multicolumn{1}{c}{contribution} \\
\multicolumn{1}{c}{$E_x$(keV)} &
\multicolumn{1}{c}{~~$J^{\pi};T_{\pi}$} &
\multicolumn{1}{c}{~~~$\alpha$~~ } &
\multicolumn{1}{c}{$S_{\alpha}$ } &
\multicolumn{1}{c}{~~$\Omega_{\alpha}^{\pi}(\%)$} &
\multicolumn{1}{c}{sum rule } &
\multicolumn{1}{c}{to $\delta_{C2}(\%)$} &
\multicolumn{1}{c}{$\sum_{\pi}S_{\alpha}^{\pi}$ } &
\multicolumn{1}{c}{ to $\delta_{C2}(\%)$} \\[1mm]
\hline
& & & & & & & & \\
0 & $\sfrac{7}{2}^-;\sfrac{1}{2}$ & $f_{7/2}$ & 2.7(11) & 0.134 &
3.33 & ~~0.45 & 3.36 & ~~0.45 \\
330 & $\sfrac{3}{2}^+;\sfrac{1}{2}$ & $d_{3/2}$ & 1.9(8)~\, & 0.157 &
2.67 & ~~0.42 & 2.45 & ~~0.39 \\
1566 & $\sfrac{1}{2}^+;\sfrac{1}{2}$ & $s_{1/2}$ & 0.7(3)~\, & 0.318 &
1.33 & ~~0.42 & 1.22 & ~~0.39 \\
4723 & $\sfrac{7}{2}^-;\sfrac{3}{2}$ & $f_{7/2}$ & 3.6(16) & 0.085 &
2.67 & $-0.11$ & 2.74 & $-0.12$ \\
4810 & $\sfrac{3}{2}^+;\sfrac{3}{2}$ & $d_{3/2}$ & 3.6(16) & 0.100 &
5.33 & $-0.27$ & 4.92 & $-0.25$ \\
5760 & $\sfrac{1}{2}^+;\sfrac{3}{2}$ & $s_{1/2}$ & 3.2(12) & 0.224 &
2.67 & $-0.30$ & 2.47 & $-0.28$ \\[2mm]
\end{tabular}
\end{ruledtabular}
\end{center}
\end{table*}

This equation provides the key to the strategy we will use in calculating
$\delta_{C2}$.  It demonstrates that there is a cancellation between the
contributions of the isospin-lesser states and
the isospin-greater states.  Moreover, if
the orbital $\alpha$ were completely full in the initial
$A$-particle wavefunction, then the Macfarlane and French sum rules \cite{FM61} 
for spectroscopic factors would require $\sum_{\pi^<} S_{\alpha}^< $ = 
$ \sfrac{1}{2} \sum_{\pi^>} S_{\alpha}^>$ and the cancellation in Eq.\,(\ref{dc2_3})
would be very strong.  In fact, the cancellation would be complete if
$\Omega_{\alpha}^< = \Omega_{\alpha}^>$.
As we will discuss further in the next section, this cancellation is not in general
complete because the radial-mismatch factors for isospin-lesser states
are larger than those for isospin-greater states.  Even so, cancellation is always
significant, and it becomes most complete when closed-shell orbitals are involved.
Furthermore, the more deeply bound the closed-shell orbital, the greater the energy
spread in the spectroscopic strength and the more complete the cancellation.  Thus, 
although the dominant contributions to $\delta_{C2}$ come from unfilled orbitals, 
we conclude that closed-shell orbitals must play a role, albeit one that decreases in
importance as the orbitals become more deeply bound. 

Based on these observations, our strategy is to use experiment to guide us in determining
which closed-shell orbitals are important enough to include.  Ideally, of course, one
would take the spectroscopic factors determined from experiment and insert them into
Eq.\,(\ref{dc2_3}) but, especially where delicate cancellations are involved, the
reliability of (forty-year-old) experimental spectroscopic factors is certainly not up to the task.  
Our strategy then is to use the shell model to calculate the spectroscopic amplitudes
in Eq.\,(\ref{dc2_1}) but to limit the sum over orbitals $\alpha$ just to those for which
large spectroscopic factors have been observed in neutron pick-up reactions.

We illustrate the strategy for the case of $^{46}$V.  The spectroscopic 
factors for neutron pick up from $^{46}$Ti have been measured
in the $(^3$He$, \alpha )$ reaction by Borlin \cite{Bo67}.  He identified
sixteen states in $^{45}$Ti, and in Table~\ref{t:Sfactor} we
record the six states with the largest spectroscopic factors, {\it i.e.}
$S > 0.5$.  We note that the errors on the experimental spectroscopic
factors are quite large, and in two cases the quoted $S_{\alpha}$ values 
(column 4) exceed the Macfarlane-French sum rule \cite{FM61} for pure
configurations (column 6).
Thus we do not use the experimental spectroscopic factor explicitly,
but take them as a guide for which orbitals should be included
in the shell-model calculation.  In the case of $^{46}$V decay, they tell us
that orbitals $f_{7/2}$, $d_{3/2}$ and $s_{1/2}$ should be included.  In column five
of Table~\ref{t:Sfactor} we give a typical value for the radial
mismatch factor, $\Omega_{\alpha}^{\pi}$, for the given orbital $\alpha$
and isospin $T_{\pi}$.  Column seven gives the contribution to
$\delta_{C2}$ from this $\alpha$ and isospin $T_{\pi}$ if the
Macfarlane-French sum rule is used for the spectroscopic factor, while
in columns eight and nine are shown the results of a detailed shell-model calculation.
The results from the Macfarlane-French sum rules and the
shell-model calculation are remarkably similar.  The summed $\delta_{C2}$ for
the shell-model calculation (the sum of all entries in column 9) is 
$0.58 \%$, nearly
a factor of two larger than our previous calculated value, which was published in
2002 \cite{TH02}.

The difference between our calculations arises as follows:  In 2002 our
shell-model calculations for $^{46}$V were based on the model space $(fp)^6$, with six 
valence nucleons occupying the $pf$-shell orbitals.  In fact, only the
$f_{7/2}$ orbital contributed importantly to the $\delta_{C2}$ 
calculation so the result was $\delta_{C2} = 0.45 - 0.12 = 0.33 \%$ (see
the two rows for the $f_{7/2}$ orbital in Table~\ref{t:Sfactor}).
Absent from this 2002 calculation was any contribution from the core orbitals,
$d_{3/2}$ and $s_{1/2}$.  In our present calculations, these orbitals are included,
with the $d_{3/2}$ orbital contributing 0.14 \% to $\delta_{C2}$ and the
$s_{1/2}$ contributing 0.11 \%.

But why stop there?  Why not include the $d_{5/2}$ and
possibly the $p$-shell orbitals in the computation?  Our answer is that
the neutron pick-up measurement saw little or no evidence for
such core states, which implies that their spectroscopic strength is distributed 
widely over many states.  In this case, the cancellation between isospin-lesser
and isospin-greater states becomes more complete and their contribution to
$\delta_{C2}$ is reduced to a level that we believe can be neglected.

With this approach, we are now in a position to revise our earlier results \cite{TH02} to
include the effects of previously ignored core orbitals.  Again using measured
spectroscopic factors from neutron pick-up reactions, we determined that
changes were required for the $A = 22$ and $26$ cases, in which $p$-shell holes
must contribute in addition to the original $sd$-shell configurations; similarly, $sd$-shell
holes were required in addition to the $pf$-shell particles for $A = 46,~50$ and $54$.
For $A = 62,~66,~70$ and $74$ in the upper $pf$-shell
there are no experimental neutron pick-up reaction measurements to guide us.
Our previously published calculations for these nuclei were based on
$(p_{3/2},~f_{5/2},~p_{1/2})^n$ model spaces using $^{56}$Ni as a closed-shell 
core.  It seemed prudent now for these cases at least to include
the $f_{7/2}$ orbital in the calculation of $\delta_{C2}$, and we have made this
change.  In the cases with
$A$ = 18 and 42, we had previously included some contribution from deeper shells;
we did not need to make any changes in the former but did add the $s_{1/2}$
and $d_{5/2}$ shells to the latter.  No additional orbitals were required for the cases with
$A = 10,~14,~30$,~34 and 38. 

\subsubsection{Radial-mismatch Factor, $\Omega_{\alpha}^{\pi}$}
\label{sss:rmf}
 
In considering the radial integrals, we benefit from a very strong
constraint: the asymptotic forms of all radial functions must match
the measured separation energies, $S_p$ and $S_n$, where $S_p$ is
the proton separation energy in the decaying nucleus and $S_n$ the
neutron separation energy in the daughter nucleus.  The basic ingredients
of these separation energies are well known and can be
found in any atomic mass tables.  It is the size of the difference
between $S_p$ and $S_n$ and the presence or absence of nodes in
the radial wave functions that are the principal factors in determining
the magnitude of $\Omega_{\alpha}^{\pi}$.

Our calculations of this mismatch factor follows the same path as that 
described in our earlier works \cite{TH02,THH77}.  We use a Saxon-Woods
potential defined for a nucleus of mass $A$ and charge $Z+1$ as:
\be
V(r) = - V_0 f(r) - V_s g(r) {\bf l}. \mbox{\boldmath$\sigma$}
+ V_C(r) - V_g g(r) - V_h h(r),
\label{VSW}
\ee
where
\bea
f(r) & = & \left \{ 1 + \exp \left ( (r-R)/a \right ) \right \}^{-1} ,  
\nonumber  \\
g(r) & = & \left ( \frac{\hbar}{m_{\pi} c} \right )^2 
\frac{1}{a_s r} \exp \left ( \frac{r - R_s}{a_s} \right )
\nonumber \\
& & ~~~~~ \times 
\left \{ 1 + \exp \left ( \frac{r - R_s}{a_s} \right )
\right \}^{-2} ,
\nonumber  \\
h(r) & = & a^2 \left ( \frac{df}{dr} \right )^2 ,
\nonumber  \\
V_C(r) & = & Z e^2 / r , ~~~~ {\rm for}~~ r \geq R_c
\nonumber  \\
 & = & \frac{Z e^2}{2 R_c} \left ( 3 - \frac{r^2}{R_c^2} \right )
 , ~~~~ {\rm for}~~ r < R_c ,
\label{Pot}
\eea
with $R = r_0 (A - 1)^{1/3}$ and
$R_s = r_s (A - 1)^{1/3}$.  
The first three terms in Eq.\,(\ref{VSW}) are the
central, spin-orbit and Coulomb terms respectively.  The fourth
and fifth terms are additional surface terms whose role we discuss
shortly.

Most of the parameters 
were fixed at standard values, $V_s = 7$ MeV, $r_s = 1.1$ fm and
$a = a_s = 0.65$ fm.  The radius of the Coulomb potential was determined
from the charge mean square radius, 
$\langle r^2 \rangle_{{\rm ch}}^{1/2}$,
of the decaying nucleus as determined from elastic electron scattering;
see Eqs.\,(21) and (22) in Ref.\,\cite{TH02}.
The well radius, $r_0$, was similarly fixed, by requiring that the
charge density constructed from the square of the proton
wave functions bound in the well should also match the charge mean
square radius.  Initially, with $V_g$ and $V_h$ set to zero, the well
depth, $V_0$, was adjusted
so that the binding energy of the least-bound orbital matched the
experimental separation energy.

From the shell model calculation, we obtained the $A$-particle wave functions,
$| i \rangle$ and $| f \rangle$, expanded into products of $(A-1)$-particle
wave functions $|\pi \rangle $ and single-particle functions
$| \alpha \rangle $.  In Eq.\,(\ref{MFpar}) and the discussion that followed it, we
noted that the radial integral should depend on the separation energies relative
to the $(A-1)$ state, $| \pi \rangle $.  We ultimately allowed this to
happen but initially we calculated the value of $\delta_{C2}$ under the
assumption that the proton and neutron radial functions, $R^p(r)$ and $R^n(r)$,
have asymptotic forms for all $\alpha $ that are fixed at the separation
energies, $S_p$ and $S_n$, to the ground state of the $(A-1)$ nucleus.  In this
case, the sums over $\pi$ can be done analytically
and the computed value of $\delta_{C2}$ becomes independent
of the shell-model effective interaction.  This result, which we label
$\delta_{C2}^{I}$, can be simply expressed with
the help of Eqs.\,(\ref{M0}) and (\ref{dc2_1}):
\be
\delta_{C2}^I \simeq 2 \Omega_{\alpha_g} .
\label{dc2I}
\ee
Here $\alpha_g$ is
the shell-model orbital of the transferred neutron in the pick-up
reaction from the $A$-particle state $|f \rangle$ to the ground state
of the $(A-1)$-particle nucleus.

We next removed our simplifying assumption and evaluated the radial
integrals with eigenfunctions of the Saxon-Woods potential whose well
depth was adjusted so that each eigenfunction matched the separation energy
of the $(A-1)$ state to which it corresponds, $| \pi \rangle $.  For an $(A-1)$ state
at excitation energy $E_x$ the corresponding separation energies are
$S_p + E_x$ and $S_n + E_x$.  We label these results $\delta_{C2}^{II}$
and note that the values now depend on the spectroscopic amplitudes, and
hence on the shell-model effective interaction, but not strongly.   

So far, we have ignored the two surface terms in Eq.\,(\ref{Pot}) by setting
$V_g = 0$ and $V_h = 0$.  It can be argued, however, that the central part of
the potential, which in principle should be determined from some Hartree-Fock
procedure, should not be continually adjusted.  Instead, any adjustments made
to match separation energies should be to the surface part of the potential rather
than to the depth of the well.  Thus, we also calculated $\delta_{C2}$ by
fixing $V_0$ separately for protons and neutrons to match the ground-state
parent separation energies, $S_p$ and $S_n$, and then adjusting the strength
of the surface term, $V_g$ (keeping $V_h = 0$) so that the asymptotic forms
matched the separation energies $S_p + E_x$ and $S_n + E_x$.
These results are labelled $\delta_{C2}^{III}$.  

Finally, our fourth method of calculation was the same as the third, 
except that it was the second surface term, $V_h$, that was adjusted to match
separation energies, keeping $V_g = 0$.  This second term, $h(r)$, is even
more strongly peaked in the surface than $g(r)$.  These results are labelled
$\delta_{C2}^{IV}$.  

On average, the method III values of $\delta_{C2}$ are about 2\% lower
than the method II values; and method IV values are about 7\% lower than
the method II values for orbitals without any radial nodes.
For orbitals with one or more nodes, there is more of the radial 
wave function in the surface region and methods III and IV produce greater
reductions.

\subsubsection{The shell-model calculations}
\label{sss:smc}

We now present our results for $\delta_{C2}$ based on the extensions
of the shell-model spaces mentioned at the end of Sect.~\ref{sss:st}.
In addition to adding the core orbitals mentioned there, however, in
some cases we have also been able to make use of more recent effective
interactions that have become available since our last work.  Specifically,
we have used the following interactions in the various mass regions of
interest:  In the $p$-shell, we use the Cohen-Kurath interactions \cite{CK65}
and the more recent PWBT interaction of Warburton and Brown \cite{WB92}.  In the
$s,d$-shell, besides the universal interaction of Wildenthal \cite{Wi84},
we employ two new versions, USD-A and USD-B, of Brown and Richter \cite{BR06}.
In the $pf$-shell we use the KB3 interaction of Kuo-Brown \cite{KB66}
as modified by Poves and Zuker \cite{PZ81}, the FPMI3 interaction of
Richter and Brown \cite{Ri91}, and the more recent GXPF1 interaction
of Honma \etal \cite{Ho02,Ho04}.  For cross-shell interactions between the 
major shells, we have used the interaction of Millener and Kurath \cite{MK75}.
It should be noted that in many cases we found it necessary to introduce
some truncations in the original model space in order to keep the calculations
tractible.

\begin{table}[t]
\begin{center}
\caption{Calculations of $\delta_{C2}$ with Saxon-Woods radial     
functions, without parentage expansions ($\delta_{C2}^I$) and
with parentage expansions
($\delta_{C2}^{II}$,
$\delta_{C2}^{III}$,
and $\delta_{C2}^{IV}$).  Note that only one sample result is shown
in each case
for $\delta_{C2}^I$, $\delta_{C2}^{II}$, $\delta_{C2}^{III}$ and
$\delta_{C2}^{IV}$, while the adopted $\delta_{C2}$ value in column 7 reflects
the results from all multiple-parentage calculations for that case; see text.
\label{t:dc2}}
\vskip 1mm
\begin{ruledtabular}
\begin{tabular}{lrrrrrrr}
& & & & & & \\[-2mm]
& \multicolumn{1}{c}{2002} & \multicolumn{5}{c}{This work}\\
\cline{3-7} \\[-3mm]
\multicolumn{1}{c}{Parent} & 
\multicolumn{1}{c}{$\delta_{C2}(\% )$} & & & & &
\multicolumn{1}{c}{$\delta_{C2}(\% )$} \\
\multicolumn{1}{c}{nucleus} &
\multicolumn{1}{c}{Ref. \protect\cite{TH02}} &
\multicolumn{1}{c}{$\delta_{C2}^{I}(\% )$} &
\multicolumn{1}{c}{$\delta_{C2}^{II}(\% )$} &
\multicolumn{1}{c}{$\delta_{C2}^{III}(\% )$} &
\multicolumn{1}{c}{$\delta_{C2}^{IV}(\% )$} &
\multicolumn{1}{c}{adopted} \\[1mm] 
\hline
& & & & & & \\
\multicolumn{2}{l}{~~~~$T_z = -1:$} & & & & & \\
$^{10}$C & 0.170(15) & 0.132 & 0.163 & 0.165 & 0.163 & 0.165(15) \\
$^{14}$O &  0.270(15) & 0.217 & 0.274 & 0.271 & 0.271 & 0.275(15) \\
$^{18}$Ne & 0.390(10) & 0.251 & 0.386 & 0.387 & 0.382 & 0.410(25) \\
$^{22}$Mg & 0.255(10) & 0.207 & 0.366 & 0.382 & 0.375 & 0.370(20) \\
$^{26}$Si & 0.330(10) & 0.223 & 0.421 & 0.407 & 0.392 & 0.405(25) \\
$^{30}$S & 0.740(20) & 0.812 & 0.714 & 0.710 & 0.713 & 0.700(20) \\
$^{34}$Ar & 0.610(40) & 0.351 & 0.680 & 0.639 & 0.579 & 0.635(55) \\
$^{38}$Ca & 0.710(50) & 0.402 & 0.840 & 0.784 & 0.702 & 0.745(70) \\
$^{42}$Ti & 0.555(40) & 0.359 & 0.881 & 0.849 & 0.780 & 0.835(75) \\[2mm]
\multicolumn{2}{l}{~~~~$T_z = 0:$} & & & & & \\
$^{26}$Al$^m$ & 0.230(10) & 0.156 & 0.292 & 0.280 & 0.271 & 0.280(15) \\
$^{34}$Cl & 0.530(30) & 0.312 & 0.583 & 0.561 & 0.498 & 0.550(45) \\
$^{38}$K$^m$ & 0.520(40) & 0.299 & 0.623 & 0.575 & 0.522 & 0.550(55) \\
$^{42}$Sc & 0.430(30) & 0.278 & 0.681 & 0.648 & 0.606 & 0.645(55) \\
$^{46}$V & 0.330(25) & 0.273 & 0.587 & 0.543 & 0.506 & 0.545(55) \\
$^{50}$Mn & 0.450(30) & 0.315 & 0.638 & 0.598 & 0.594 & 0.610(50) \\
$^{54}$Co & 0.570(40) & 0.376 & 0.760 & 0.688 & 0.706 & 0.720(60) \\
$^{62}$Ga & 1.05(15) & 1.31 & 1.22 & 1.19 & 1.14 & 1.20(20) \\
$^{66}$As & 1.15(15) & 1.32 & 1.41 & 1.34 & 1.24 & 1.35(40) \\
$^{70}$Br & 1.00(20) & 1.43 & 1.41 & 1.31 & 1.10 & 1.25(25) \\
$^{74}$Rb & 1.30(40) & 1.68 & 1.60 & 1.47 & 1.12 & 1.50(30) \\[2mm]

\end{tabular}
\end{ruledtabular}
\end{center}
\end{table}

We made calculations for all twenty superallowed transitions considered in
our earlier work \cite{TH02,HT05}, and for each we calculated $\delta_{C2}$
in the four methods, I-IV, described in Sect.\,\ref{sss:rmf} and with the
several interactions listed in the previous paragraph. In Table~\ref{t:dc2}
we record only {\em one sample result} for $\delta_{C2}^{I}$, $\delta_{C2}^{II}$,
$\delta_{C2}^{III}$ and $\delta_{C2}^{IV}$ for each nucleus listed.  However, 
our ``adopted $\delta_{C2}$" values result from our assessment 
of {\em all} multiple-parentage calculations made for each decay,
not just those shown in the previous three columns.  The uncertainty 
assigned to each adopted value reflects the uncertainty in the radius of 
the Saxon-Woods potential (resulting from an uncertainty in the nuclear rms
radius to which it is adjusted), the spread of results obtained with
different shell-model interactions, and the spread of results 
obtained with the different procedures labelled II, III and IV in the Table.
 
\subsection{ Isospin-Mixing Correction, $\delta_{C1}$}
\label{ss:dc1}

The second (and smaller) contribution to $\delta_C$ is the
isospin-mixing correction, $\delta_{C1}$.  For its evaluation, the radial integrals
are all set to unity, but the spectroscopic amplitudes in 
Eq.\,(\ref{MFpar}) are not required to satisfy hermiticity.  Calculations
of this correction turn out to be very sensitive to the details of
the shell-model computation.  This would be a very unfortunate
property if we were not able to adopt certain strategies that act
to reduce the model dependence considerably.

\begin{table*}[t]
\begin{center}
\caption{Shell-model calculations of the                       
isospin-mixing correction, $\delta_{C1}$.
\label{t:dc1new}}
\vskip 1mm
\begin{tabular}{lcccccccc}
\hline
\hline \\[-3mm]
& & & & & & & & \\[-2mm]
& \multicolumn{2}{c}{Measured} & 2002 & \multicolumn{5}{c}{This work} \\
\cline{5-9} \\[-3mm]
Parent  & \multicolumn{2}{c}{IMME coefficients \protect\cite{Br98}}
& \multicolumn{1}{c}{$\delta_{C1}(\%)$}
& \multicolumn{1}{c}{$E_x(0^{+})$}
& \multicolumn{1}{c}{$E_x(0^{+})$}
& \multicolumn{1}{c}{$\delta_{C1}(\%)$}
& \multicolumn{1}{c}{$\delta_{C1}(\%)$}
& \multicolumn{1}{c}{$\delta_{C1}(\%)$} \\
\cline{2-3} \\[-3mm]
nucleus  & \multicolumn{1}{c}{b (keV)}
& \multicolumn{1}{c}{c (keV)}
& \multicolumn{1}{c}{Ref. \protect\cite{TH02}}
& \multicolumn{1}{c}{expt}
& \multicolumn{1}{c}{SM}
& \multicolumn{1}{c}{unscaled}
& \multicolumn{1}{c}{scaled}
& \multicolumn{1}{c}{adopted} \\[1mm]
\hline
& & & & & & & & \\[-3mm]
~~~~$T_z = -1$: & & & & & & & & \\
   $^{10}$C     & $  -1.546 $ & $   0.362 $ & $ 0.010(10) $
   & $  6.18   $ & 9.24  & $  0.005  $ & $   0.011 $ & $ 0.010(10) $  \\
   $^{14}$O    & $  -2.493 $ & $   0.337 $ & $ 0.050(20) $
   & $  6.59   $ &  6.64   & $   0.049 $ & $   0.050 $ & $0.055(20)$ \\
   $^{18}$Ne      & $  -3.045(1) $ & $ 0.347(1)$ & $ 0.230(30) $
   & $   3.71  $ & 4.07  & $  0.116  $ & $    0.140$ & $0.155(30)$\\
   $^{22}$Mg      & $ -3.814(1) $ & $ 0.315(1)$ & $ 0.010(10) $
   & $  6.24   $ & 6.21   & $  0.010  $ & $  0.010  $ & $ 0.010(10) $ \\
   $^{26}$Si     & $ -4.535(2) $ & $ 0.302(2)$ & $ 0.040(10) $
   & $  3.59   $ & 3.86   & $  0.022  $ & $   0.026 $ & $ 0.030(10) $ \\
   $^{30}$S      & $ -5.185(2) $ & $ 0.275(2)$ & $ 0.195(30) $
   & $  3.79   $ & 3.80   & $  0.137    $ & $ 0.138   $ & $ 0.155(20) $ \\
   $^{34}$Ar      & $  -5.777(2) $ & $ 0.286(2)$ & $ 0.030(10) $
   & $  3.92   $ & 3.97   & $  0.023  $ & $  0.023  $ & $ 0.030(10) $ \\
   $^{38}$Ca     & $  -6.328(3) $ & $ 0.284(3)$ & $ 0.020(10) $
   & $  3.38   $ & 3.21   & $  0.026  $ & $  0.023  $ & $ 0.020(10) $ \\
   $^{42}$Ti    & $ -6.712(3) $ & $ 0.287(3)$ & $ 0.220(100) $
   & $  1.84   $ & 3.16   & $ 0.038   $ & $ 0.114   $ & $ 0.100(20) $ \\[5mm]
~~~~$T_z = 0 $: & & & & & & & \\
   $^{26}$Al$^m$ & $ -4.535(2) $ & $ 0.302(2)$ & $ 0.040(10) $
   & $ 3.59    $ & 3.86   & $ 0.025   $ & $ 0.028   $ & $ 0.030(10) $ \\
   $^{34}$Cl     & $ -5.777(2) $ & $ 0.286(2)$ & $  0.105(20) $
   & $  3.92   $ & 3.97   & $ 0.091   $ & $ 0.093   $ & $0.100(10)$ \\
   $^{38}$K$^m$ & $ -6.328(3)$& $ 0.284(3)$ & $ 0.100(20) $
   & $  3.38   $ & 3.21   & $  0.099   $ & $ 0.089   $ & $ 0.105(20) $ \\
   $^{42}$Sc     & $ -6.712(3) $ & $ 0.287(3)$ & $ 0.060(30) $
   & $  3.30^a $ & 5.05   & $ 0.007   $ & $ 0.017   $ & $ 0.020(10) $ \\
   $^{46}$V   & $-7.327(10)$ & $0.276(11)$ & $ 0.095(20) $
   & $ 3.57^a  $ & 4.86   & $ 0.040   $ & $ 0.075   $ & $ 0.075(30) $ \\
   $^{50}$Mn   & $-7.892(30)$ & $0.259(30)$ & $ 0.055(20) $
   & $ 3.69    $ & 3.62   & $ 0.057    $ & $ 0.054    $ & $ 0.045(20) $ \\
   $^{54}$Co   & $-8.519(25)$ & $0.276(25)$ & $ 0.040(15) $
   & $ 2.56    $ & 2.26   & $0.058    $ & $0.045    $ & $ 0.050(30) $ \\
   $^{62}$Ga    & $ -9.463(70) $ & $ 0.265(25)^b$  & $ 0.330(40) $
   & $ 2.33    $ & 2.32   & $ 0.221   $ & $ 0.219   $ & $ 0.275(55) $ \\
   $^{66}$As     & $ -9.95(15)  $ & $ 0.262(25)^b$ & $ 0.250(40) $
   & $ 2.17^c$ & 1.89   & $ 0.210   $ & $ 0.159   $ & $0.205(45) $ \\
   $^{70}$Br     & $ -10.48(23) $ & $ 0.260(25)^b$ & $ 0.350(40) $
   & $ 2.01    $ & 2.05   & $ 0.332   $ & $ 0.346   $ & $ 0.350(40) $ \\
   $^{74}$Rb     & $ -10.82(25) $ & $ 0.258(25)^b$ & $ 0.130(60) $
   & $ 0.508   $ & 0.523  & $ 0.122    $ & $ 0.129    $ & $ 0.130(60)^d $ \\[2mm]
\hline
\hline
\\
\multicolumn{9}{l}{$^a$ \footnotesize Second excited $0^+$ state; shell-model calculations
indicate this state takes up most of the depletion from the analog state.} \\
\multicolumn{9}{l}{$^b$ \footnotesize Estimated: extrapolated from a fit to $c$ coefficients
in $0^+$ states in $A = 4n + 2$ nuclei, $10 \leq A \leq 58$; the data were taken from Ref.
\protect\cite{Br98}.} \\
\multicolumn{9}{l}{$^c$ \footnotesize Estimated: value is the average of the excitation energy of the
$0^+$ states in $^{62}$Zn and $^{70}$Se.} \\
\multicolumn{9}{l}{$^d$ \footnotesize No new calculations were performed for $^{74}$Rb.}
\end{tabular}
\end{center}
\end{table*}

There are three ways in which we incorporated charge dependence in our shell-model
calculation.  First, the single-particle energies of the proton
orbits were shifted relative to those of the neutrons.  The amount of
shift was determined from the spectrum of single-particle states
in the closed-shell-plus-proton versus the closed-shell-plus-neutron
nucleus, where the closed shell was taken to be the nucleus used
as a closed-shell core in the shell-model calculation.
We took these single-particle shifts 
from experiment and did not adjust them.

Second, we added a two-body Coulomb interaction among the valence protons
and adjusted its strength so that the measured $b$-coefficient of the
isobaric multiplet mass equation (IMME) was
exactly reproduced. Third, we introduced a charge-dependent nuclear
interaction by increasing all the $T = 1$ proton-neutron matrix
elements by about $2\%$ relative to the neutron-neutron matrix elements.  
The precise amount of this increment was determined by requiring agreement
with the measured $c$-coefficient of the IMME.  This strategy
of constraining the charge-dependence in the effective interaction
by requiring it to reproduce the coefficients of the IMME was adopted
from the work of Ormand and Brown\cite{OB85,OB89}.

Experimental data were used in one more way to constrain our
calculations.  If isospin were an exact symmetry, then the parent $0^+$ ($T=1$) state
would decay exclusively to its analog state in the daughter nucleus.  Beta transitions 
to all other $0^+$ states in the daughter would be strictly forbidden.
But, with isospin symmetry broken, weak transitions (with branching ratios
measured in parts per million) can occur to these other $0^+$
states.  In this case, we write the Fermi matrix element squared to the $n^{th}$
non-analog $0^+$ state as
\be 
|M_F^n|^2 = 2 \delta_{C1}^n
\label{MFn}
\ee
and the reduction in the analog transition Fermi
matrix element squared as
\be
|M_F|^2 = 2 ( 1 - \delta_{C1}) ,
\label{MF0}
\ee
neglecting, in this context, the contribution of $\delta_{C2}$.
If all the $0^+$ states of a given model space had the same $T = 1$ isospin
designation, then the effect of isospin-symmetry breaking terms in the
Hamiltonian would be to deplete the analog-transition strength by an
amount that is exactly matched by the sum of the strengths to the
non-analog states: {\it i.e.}
\be
\delta_{C1} \simeq \sum_n \delta_{C1}^n .
\label{dc1sum}
\ee
In practice, with large shell-model calculations the $0^+$ states in
the model space will include some states whose isospin designation
is not $T = 1$; and
Eq.\,(\ref{dc1sum}) is not then exactly correct.  Nevertheless, it remains
approximately true.

Significantly, in many cases the bulk of the analog state depletion
shows up in a single excited $0^+$ state, usually (but not always) the
first excited one.  This allows us once again to use experiment to
constrain and refine our calculation.
In the limit of only two-state mixing, perturbation theory
would indicate that
\be
\delta_{C1} \propto \frac{1}{(\Delta E)^2}
\label{propDE2}
\ee
where $\Delta E$ is the energy separation of the analog and
non-analog $0^+$ states.  Again, this is not an exact result,
but it does highlight the importance of the shell-model Hamiltonian
producing a good quality spectrum of $0^+$ states with, in particular,
the first excited non-analog $0^+$ state calculated to have an excitation 
energy close to its experimental value\footnote{In a few cases, the state
calculated to have the largest charge-dependent admixture was the second
excited $0^+$ state.  In these cases we optimized the agreement between
theory and experiment for the excitation energy of that state}.  This is not always
possible to achieve in the shell model, especially near closed shells 
where excited $0^+$ states tend to exhibit strong deformations.
We used two strategies to bring the calculation into line with
experimental information.  Our first was to adjust the
centroids of the shell-model Hamiltonian matrix elements specifically
to get the excited $0^+$ state at about the right energy.  Our second
was to scale our calculated $\delta_{C1}$ value by a factor
$(\Delta E)_{\rm theo}^2 / (\Delta E)_{\rm expt}^2$, the ratio of the
square of the excitation energy of the first excited $0^+$ state
in the model calculation to that known experimentally.

We list in Table~\ref{t:dc1new} the experimental values \cite{Br98}
of the IMME coefficients, $b$ and $c$, and the known excitation
energy $E_x(0^+)$ of the first (or second) excited $0^+$ state in the
daughter nuclei.  As explained, all our shell-model
calculations were adjusted to reproduce exactly the values
of $b$ and $c$, and to match, as closely as possible the excitation
energy of the excited $0^+$ state.  We compensated for any remaining
discrepancies between the calculated and experimental values of
$E_x(0^+)$ by scaling the results for $\delta_{C1}$.  As in
Table~\ref{t:dc2}, we give (in columns 6--8) the results from
{\em one sample calculation} for each nucleus.  Then in column
nine we present adopted $\delta_{C1}$ values that result from
our assessment of the results of {\em all} calculations made for
each decay, not just the ones shown in columns 6--8; the
uncertainties were chosen to encompass the spread in the results
from those calculations and to include the uncertainty in the
IMME $b$ and $c$ coefficients.  For comparison, in column 4 we
list the values we adopted for $\delta_{C1}$ in 2002\cite{TH02}.  
Our strategies have remained unchanged, but here we have 
additionally used some more recent shell-model effective
interactions as listed in Sect.~\ref{sss:smc}.  In nearly all cases,
the new values of $\delta_{C1}$ agree with the old values within their 
stated uncertainties.

\begin{table}[t]
\begin{center}
\caption{Shell-model calculations of $\delta_{C1}^1$ for Fermi decay to
the first excited $0^+$ state; see Eq.\,(\protect\ref{MFn}).  The results are compared with
experimental measurements where they are known.  All values are expressed in \%.                        
\label{t:dc1exnew}}
\vskip 1mm
\begin{ruledtabular}
\begin{tabular}{lccccc}
& & & & & \\[-3mm]
Parent & 2002 & \multicolumn{3}{c}{This work} & \\
\cline{3-5} \\[-2mm]
nucleus
& \multicolumn{1}{c}{value\protect\cite{TH02}}  
& \multicolumn{1}{c}{unscaled}
& \multicolumn{1}{c}{scaled}
& \multicolumn{1}{c}{Adopted} 
& \multicolumn{1}{c}{expt }  \\[1mm]
\hline
& & & & & \\[-3mm]
\multicolumn{2}{l}{~~~~$T_z = 0 $:} & & & & \\
   $^{38}$K$^m$ & 
   $ 0.090(30) $ &$  0.068$ & $ 0.062$ & $0.085(30)$ &  
   $ <0.28\footnotemark[1] $  \\
   $^{42}$Sc    & 
   $ 0.020(20)  $ &$ 0.007   $ & $ 0.027   $ & $0.015(15)$ &  
   $ 0.040(9)\footnotemark[2] $ \\
   $^{46}$V   & 
   $ 0.035(15) $ & $ 0.008$ & $ 0.024$ & $ 0.025(20)$ & 
   $ 0.053(5)\footnotemark[1] $ \\
   $^{50}$Mn   & 
   $ 0.045(20) $ &$ 0.049$ & $ 0.047 $ & $ 0.040(20) $ &  
   $ <0.016\footnotemark[1] $ \\
   $^{54}$Co  & 
   $ 0.040(20) $ & $0.049$ & $0.038 $ & $ 0.050(20) $ & 
   $ 0.035(5)\footnotemark[1] $ \\
   $^{62}$Ga    & 
   $0.085(20) $ & $  0.160$ & $ 0.159$  & $ 0.120(40) $ &  
   $ \leq 0.040(15)\footnotemark[3] $ \\
   $^{66}$As    & 
   $0.020(20) $ &  $  0.110$ & $ 0.087$ & $ 0.050(30) $ &  \\
   $^{70}$Br    & 
   $0.070(20) $ &  $  0.226$ & $ 0.235$ & $ 0.150(80) $ &  \\
   $^{74}$Rb    & 
   $0.050(30) $ &  $  0.045$& $ 0.047$ & 0.050(30)\footnotemark[4] &  
    $ \leq 0.075\footnotemark[5] $ \\
\end{tabular}
\end{ruledtabular}
\footnotetext[1]{From Hagberg \protect\etal (1994)  
\protect\cite{Ha94}  }
\footnotetext[2]{From Daehnick and Rosa (1985) \protect\cite{DR85} averaged
with earlier results.  }  
\footnotetext[3]{From Hyland \protect\etal (2006)  
\protect\cite{Hy06}  }
\footnotetext[4]{No new calculations were performed for $^{74}$Rb.}
\footnotetext[5]{From Piechaczek \protect\etal (2003)  
\protect\cite{Pi03}  }
\end{center}
\end{table}

For the heavier nuclei there are experimental data on Fermi
transitions to the non-analog excited $0^+$ states.  The
measured branching ratios \cite{Hy06,Ha94,DR85,Pi03} have been converted to $\delta_{C1}^1$
values, {\it via} Eq.\,(\ref{MFn}), and listed in Table~\ref{t:dc1exnew}.
Again, for each nucleus, we list just one representative
calculation and our adopted value.  The assigned error reflects
both the spread among the different calculations and the uncertainties 
in the IMME coefficients.  Our 2002 adopted values \cite{TH02}
are also listed.  For nuclei $38 \leq A \leq 54$, with the possible
exception of $^{50}$Mn, the agreement between theory and experiment 
is entirely satisfactory.  But in the upper $pf$-shell, 
the calculated value for $^{62}$Ga is three times larger
than measured in recent experiments \cite{Hy06}.  Shell-model
calculations in this region are complicated by the massive size of
the Hamiltonian matrices.  To keep our calculations tractible, we
kept the $f_{7/2}$ shell closed in these cases, but there
is considerable evidence \cite{Ho04} that this could be a
poor assumption.

\section{The Radiative Correction}
\label{s:radc}
\subsection{Prior to 1990}
\label{ss:pre90}

Conventionally, the radiative correction has been separated into two parts, one that
contains the nucleus-dependent terms, called the `outer' radiative correction, and
one that is independent of the nucleus, the `inner' radiative correction. Principally
due to the work of Marciano and Sirlin (for example, Refs.\,\cite{Si67,MS84,MS86}), the
radiative correction applied to the uncorrected $\beta$-decay rate $\Gamma_{\beta}^0$ 
was expressed as follows:
\bea
\Gamma_{\beta} & = & \Gamma_{\beta}^0 (1 + \delta_R^{\prime}) (1 + \DRV )
\label{rate} \\[2mm]
\delta_R^{\prime} & = & \frac{\alpha}{2 \pi} \left [ \overline{g}(E_m)+\delta_2+\delta_3 \right ] ~~
\stackrel{large~E_m}{\longrightarrow}
\nonumber \\
& \longrightarrow & \frac{\alpha}{2 \pi}
\left [ 3 \ln \left ( \frac{m_p}{2 E_m} \right ) + \frac{81}{10}
- \frac{4 \pi^2}{3} +\delta_2+\delta_3 \right ] ~~~~
\label{dr1} \\
\DRV & = & \frac{\alpha}{2 \pi} \left [ 3 \ln \frac{\mW}{m_p}
+ \ln \frac{\mW}{\mA} + 2C  \right .
\nonumber \\*
& & ~~~~~~~~~~~~~~ \left . - 4 \ln \frac{\mW}{\mZ} + \A_g \right ]
\label{DR1} \\
& = & \frac{\alpha}{2 \pi} \left [ 4 \ln \frac{\mZ}{m_p} +
\ln \frac{m_p}{\mA} +2C +\A_g \right ],
\label{DR2}
\eea
where $E_m$ is the maximum electron energy in $\beta$-decay, and $\mW$,
$m_p$, $\mZ$ are the masses of the $W$-boson, proton and $Z$-boson.  The
separation into outer and inner terms is accommodated in $\delta_R^{\prime}$
and $\DRV$ respectively.

In the outer correction, $\delta_R^{\prime}$, the order-$\alpha$ term
contains the function $\overline{g}(E_m)$: it is the average over the
beta energy spectrum of the function $g(E, E_m )$, which was defined by
Sirlin (see Eq.\,(20b) of Ref.\,\cite{Si67}) and is not reproduced here.
Its large-$E_m$ limit is shown in Eq.\,(\ref{dr1}), indicating that the
expression is dominated by the logarithm, $\ln (m_p /(2 E_m))$.  The
last two terms in the outer correction, $\delta_2$ and $\delta_3$, 
represent corrections to order $Z\alpha^2$ and $Z^2\alpha^3$ respectively.
The origin of the $\overline{g}(E_m)$ term -- together with that of the leading term
in the inner radiative correction, $3 \ln (\mW / m_p )$ -- is the $\gamma W$-box
and bremsstrahlung diagrams, which are taken together to remove the divergence
as the photon energy goes to zero.  Both $\delta_2$ and $\delta_3$ also
come from a standard QED calculation of the $\gamma W$-box and bremsstrahlung
graphs \cite{Si87,JR87}, but in their case the electron was allowed to interact
with the Coulomb field of the nucleus.  Care was taken not to double count with
the Fermi function.  The calculation was complete to order $Z \alpha^2$ but
only estimated in order $Z^2 \alpha^3$. 

In the inner correction, $\DRV$, the second and third terms, $\ln (\mW / \mA )
+ 2C$, like the first term, also represent a $\gamma W$ box graph, but this time
it involves an axial-vector weak interaction.  The evaluation of this graph can
be divided into two energy regimes:  the high-energy (or short-distance) part given
by the logarithm, and the low-energy (or long-distance) part denoted by
$2C$.  The parameter $\mA$, referred to as the low-energy cut-off, divides
these two energy regimes.  Marciano and Sirlin \cite{MS84} allowed it to
take on a range of values, 400 MeV $\leq \mA \leq$ 1600 MeV (revised slightly by
Sirlin \cite{Si94} to be $m_{a_1}/2 \leq \mA \leq 2 m_{a_1}$, with
$m_{a_1}$ being the $A_1$-vector-meson mass).  The low-energy component,
$2C$, was approximated by its Born contribution
\be
C \rightarrow C_{\rm Born} = 3 \gA (0.266) (\mu_p + \mu_n ) = 0.885 ,
\label{CBorn1}
\ee
where $\gA = 1.26$ is the axial vector coupling constant accepted at the time
and $(\mu_p + \mu_n ) = 0.88$ is the nucleon isoscalar magnetic moment.  The
factor $0.266$ is the value of the loop integral that was rendered
finite by the use of dipole form factors for the nucleon
electromagnetic, $\gamma N$, and axial-vector, $W N$, vertices.
The fourth term in Eq.\,(\ref{DR1}), with the logarithm $\ln (\mW / \mZ )$,
arises from $Z W$-box graphs; while the last term,
$\A_g$, represents a small perturbative QCD correction that was 
evaluated by Marciano and Sirlin \cite{MS86} to be $\A_g = -0.34$.

The value of the outer radiative correction as defined in Eq.\,(\ref{dr1}), 
ranges from 1.39-1.65\% for the known superallowed emitters (see
Ref.\,\cite{TH02}).  Following Sirlin \cite{Si87}, the assigned uncertainties
are set equal to $(\alpha /2 \pi ) \delta_3$ as an estimate of the error
made in stopping the calculation at that order.  The value of the inner
radiative correction as obtained from Eq.\,(\ref{DR2}) with $C$ from
Eq.\,(\ref{CBorn1}) is
\cite{MS86,Si94} 
\be
\DRV (old) = 2.40(8) \% .
\label{DRVvalu1}
\ee

These results provide the essential foundation of the radiation corrections
still used today.  However a number of improvements have been introduced in the
intervening 17 years.

\subsection{A nuclear-structure dependent term}
\label{ss:nsd}

The low-energy part of the $\gamma W$-box diagram for an axial-vector
weak interaction, denoted $2C$, was approximated by its Born
contribution in Eq.\,(\ref{CBorn1}), and was evaluated on a single nucleon.
However, in a finite nucleus with many nucleons present, Jaus and Rasche \cite{JR90}
observed that the two hadronic-interaction vertices, $\gamma N$ and
$W N$, do not have to be with the same nucleon.  Thus, in finite
nuclei there can be two types of contributions: those in which
$\gamma N$ and $W N$ vertices are with the same nucleon and those in
which they are not.  The evaluation of the former terms 
yields expressions \cite{To92} that are proportional
to $\tau_+$, the isospin ladder operator, and so are also proportional
to the Fermi $\beta$-decay operator.  Therefore, they
produce a universal correction -- the same in all nuclei -- with
the value $C_{\rm Born}$, which is given in Eq.\,(\ref{CBorn1}).  The remaining
terms, those in which the interactions are with different nucleons, must be evaluated
with two-body operators that depend on the nuclear structure of the states involved.
Thus, the expression for $C$ given in Eq.\,(\ref{CBorn1}) must be replaced by the
following equation:
\be
C = C_{\rm Born} + C_{NS},
\label{CNS}
\ee
where $C_{NS}$ comprises the nuclear-structure dependent terms.  Calculations of $C_{NS}$
were first made in 1992 \cite{To92,BBJR92}.

A further modification was introduced in 1994 \cite{To94}.  In calculations
of $C_{\rm Born}$ that had been made up to that time, the
axial-vector and electromagnetic coupling constants, $\gA$ and
$(\mu_p + \mu_n )$ -- see Eq.\,(\ref{CBorn1}) -- had been given their free-nucleon
values.  Yet there is ample evidence in nuclear physics that coupling 
constants for spin-flip processes are quenched in the nuclear
medium, with the amount of quenching varying from nucleus to nucleus.  Thus,
one should really be replacing $C_{\rm Born}^{\rm free}$, the value
obtained with free-nucleon coupling constants, with $C_{\rm Born}^{\rm quenched}$.
However, to separate the nucleus-dependent and nucleus-independent parts of the
latter, we write
\bea
C_{\rm Born}^{\rm quenched} & = & q C_{\rm Born}^{\rm free}
\nonumber \\
& = & C_{\rm Born}^{\rm free} + ( q-1 ) C_{\rm Born}^{\rm free}
\label{CBornq}
\eea
where $q$ is the factor by which the product of the weak and
electromagnetic coupling constants is reduced in the medium relative
to its free-nucleon value.

The first term in Eq.\,(\ref{CBornq}), which remains universal, is
retained in the inner radiative correction, replacing $C$ in Eq.\,(\ref{DR1}).
The second term becomes part of a separate nuclear-structure-dependent
radiative correction, $\delta_{NS}$, which also includes $C _{NS}^{\rm quenched}$,
the value of $C _{NS}$ recalculated with quenched operators.  This
correction is written as
\be 
\delta_{NS} = \frac{\alpha}{\pi} \left [ C_{NS}^{\rm quenched} +
(q-1) C_{\rm Born}^{\rm free} \right ],
\label{dns}
\ee
and is incorporated with the other nuclear-structure-dependent
correction term, $\delta_C$ -- see Eq.\,(\ref{Ftfactor}).
Calculated values of $\delta_{NS}$ \cite{TH02} range from -0.360\% to +0.030\%,
each generally being smaller in magnitude than the corresponding value of $\delta_C$.

We return to $\delta_{NS}$ in Sect.~\ref{ss:col}.

\subsection{Improvements to $\DRV$ }
\label{ss:oar}

In 2005, Czarnecki, Marciano and Sirlin \cite{CMS05} revisited the
$\O (\alpha^2)$ correction for neutron beta decay.  They began by
trivially updating the value of $C_{\rm Born}^{\rm free}$ to reflect
the current value of the axial-vector coupling constant, $\gA = 1.27$,
to get
\be
C_{\rm Born}^{\rm free} = 0.891,
\label{CBorn3}
\ee
which replaces the value given in Eq.\,(\ref{CBorn1}).

They then went on to re-evaluate $\DRV$, focusing particularly on the
leading log corrections.  Using an established renormalization group
summation \cite{MS86} for the leading short-distance logs, $S(m_p,m_Z)$,
they extended the method to the lower energy region between $2E_m$ and
$m_p$ to obtain $L(2E_m,m_p)$.  This resulted in the replacements
\bea
1 + \frac{2\alpha}{\pi} \ln \frac{m_Z}{m_p} & \rightarrow & S(m_p , \mZ ) = 1.02248 \\
1 + \frac{3\alpha}{2\pi} \ln \frac{m_p}{2E_m} & \rightarrow &  L(2 E_m , m_p ) ,
\label{SL}
\eea
where
\be
L(2 E_m , m_p ) = 1.026725 \left [ 1 - \frac{2 \alpha (m_e)}{3 \pi }
\ln \frac{2 E_m}{m_e} \right ]^{9/4} .
\label{LEM}
\ee
The complete radiative correction, RC, including order $Z \alpha^2$ and
$Z^2 \alpha^3$ terms, could then be written \cite{CMS05}
\bea
1 + RC & = & \left \{ 1 + \frac{\alpha }{2 \pi} \left [ \overline{g}(E_m)
- 3 \ln \frac{m_p}{2 E_m}\right ] \right \} 
\nonumber \\
&  \times & \! \! \! \left \{
L(2 E_m,m_p) + \frac{\alpha}{2 \pi} \left [ 2 C_{\rm Born}^{\rm free}
+ \delta_2 + \delta_3 \right ] \right \}
\nonumber \\
&  \times & \! \! \! \left \{ S(m_p, \mZ) + \frac{\alpha (m_p)}{2 \pi}
\left [ \ln \frac{m_p}{\mA} + \A_g \right ] + NLL \right \},
\nonumber \\
& &
\label{radc4}
\eea
where $NLL$ is a next-to-leading log correction that Czarnecki
{\it et al.}~estimate to be $NLL = -0.0001$.  The coefficient $\alpha (m)$
is a running QED coupling constant whose value at $m = m_p$ is $1/133.986$ 
and at $m = m_e$ is $1/137.089$ \cite{CMS05}.

This new result can still be organized to preserve the separation of
nucleus-dependent and nucleus-independent components. The separation we
hereby adopt is
\bea
1 + \delta_R^{\prime} & = & 
\left \{ 1 + \frac{\alpha }{2 \pi} \left [ \overline{g}(E_m)
- 3 \ln \frac{m_p}{2 E_m}\right ] \right \} 
\nonumber \\
& & \times \left \{
L(2 E_m,m_p) + \frac{\alpha}{2 \pi} \left [ 
\delta_2 + \delta_3 \right ] \right \}
\label{dr3}
\eea
\bea
1 + \DRV & = &
S(m_p, \mZ) + \frac{\alpha}{\pi} C_{\rm Born}^{\rm free}
\nonumber \\
& & 
+ \frac{\alpha (m_p)}{2 \pi}
\left [ \ln \frac{m_p}{\mA} + \A_g \right ] + NLL. 
\label{DR4}
\eea
We will use this separation here and in our future work on superallowed $\beta$     
decay.  It results in a small change to the values of
$\delta_R^{\prime}$ and $\DRV $ that we used in our recent review \cite{HT05}.

\subsection{Reduced uncertainty for $\DRV$}
\label{ss:eeDRV}

In Sect.\,\ref{ss:pre90} we explained that the terms $\ln (\mW / \mA) + 2C$ in
Eq.\,(\ref{DR1}) arose from the $\gamma W$-box graph for an axial-vector
weak interaction.  These two terms came from splitting the evaluation of
this graph into two energy regimes.  The division between the two regimes
was chosen to be $\mA = 1.2$ GeV \cite{Si94}, roughly the mass of the $A_1$
resonance, and its range of uncertainty was taken to be from $\mA/2$ to
$2\mA$.  This $ad~hoc$ range determination actually produced the largest
single contributor to the uncertainty in the CKM matrix element, $V_{ud}$.

To reduce the hadronic uncertainty in the radiative correction,
Marciano and Sirlin \cite{MS06} have looked again at the $\gamma W$-box
graph for an axial-vector weak interaction.  This time they split it into
three energy regimes, rather than two and, where possible, they drew on
independent information to control their results:
\bi

\item {\it Short distances, $(1.5~{\rm GeV})^2 \leq Q^2 < \infty$}:  This is
a domain where QCD corrections remain perturbative.  Marciano and Sirlin added higher-order
terms, noting that these terms are identical (in the chiral limit) to QCD
corrections to the Bjorken sum rule for polarized electroproduction and can
therefore be obtained from well-studied calculations for that process.

\item {\it Intermediate distances, $(0.823~{\rm GeV})^2 \leq Q^2
< (1.5~{\rm GeV})^2$}:  In this region, they used an interpolation function
between low and high energies, motivated by vector-meson and axial-vector-meson
dominance.  By limiting the number of terms to three, they had sufficient matching
conditions to determine the coefficients uniquely.

\begin{table} [t]
\begin{center}
\caption{Calculated transition-dependent radiative
correction,
$\delta_R^{\prime}$, in percent units, and the component
contributions.  In our previous works ($eg.$ Ref.\,\cite{TH02})
$\delta_R^{\prime}$ was defined as the sum of the contents of
columns 2-4; this result is given in column 5 and labeled ``Former
$\delta_R^{\prime}$."  As explained in the text, we have now redefined
$\delta_R^{\prime}$ to include the additional term in column 6; the
new values for $\delta_R^{\prime}$ are given in the last column.
\label{t:tab1} }
\vskip 1mm
\begin{ruledtabular}
\begin{tabular}{lcccccc}
& & & & & &  \\[-2mm]
Parent & & & & Former & & Redefined \\
mucleus
& $\frac{\alpha}{2\pi } \overline{g} (E_m)$ 
& $\frac{\alpha}{2\pi } \delta_2$ 
& $\frac{\alpha}{2\pi } \delta_3$
& $\delta_R^{\prime}$ 
& $\frac{\alpha}{2\pi } \delta_{\alpha^2}$
& $\delta_R^{\prime}$ \\[1mm]
\hline 
& & & & & &  \\[-2mm]
\multicolumn{2}{l}{~~~~$T_z = -1$:} & & & & & \\
   $^{10}$C  & 1.468  & 0.180 &  0.004 & 1.652 & 0.027 & 1.679(4)  \\ 
   $^{14}$O  & 1.286  & 0.226 &  0.008 & 1.520 & 0.023 & 1.543(8)  \\   
   $^{18}$Ne & 1.204  & 0.268 &  0.012 & 1.484 & 0.022 & 1.506(12)  \\ 
   $^{22}$Mg & 1.122  & 0.307 &  0.017 & 1.446 & 0.020 & 1.466(17)  \\ 
   $^{26}$Si & 1.055  & 0.342 &  0.023 & 1.420 & 0.019 & 1.439(23)  \\ 
   $^{30}$S  & 1.005  & 0.371 &  0.029 & 1.405 & 0.018 & 1.423(29)  \\ 
   $^{34}$Ar & 0.963  & 0.396 &  0.035 & 1.395 & 0.017 & 1.412(35)  \\ 
   $^{38}$Ca & 0.929  & 0.426 &  0.042 & 1.397 & 0.017 & 1.414(42)  \\ 
   $^{42}$Ti & 0.906  & 0.456 &  0.050 & 1.412 & 0.016 & 1.428(50)  \\[2mm] 
\multicolumn{2}{l}{~~~~$T_z = 0$:} & & & & & \\
   $^{26}$Al$^m$ & 1.110 & 0.328 &  0.020 & 1.458 & 0.020 & 1.478(20)  \\ 
   $^{34}$Cl & 1.002  & 0.390 &  0.032 & 1.425 & 0.018 & 1.443(32)  \\ 
   $^{38}$K$^m$ & 0.964  & 0.420 &  0.039 & 1.423 & 0.017 & 1.440(39)  \\ 
   $^{42}$Sc & 0.939  & 0.451 &  0.047 & 1.436 & 0.017 & 1.453(47)  \\ 
   $^{46}$V  & 0.903  & 0.472 &  0.054 & 1.429 & 0.016 & 1.445(54)  \\ 
   $^{50}$Mn & 0.873  & 0.494 &  0.062 & 1.430 & 0.015 & 1.445(62)  \\ 
   $^{54}$Co & 0.844  & 0.513 &  0.071 & 1.428 & 0.015 & 1.443(71)  \\ 
   $^{62}$Ga & 0.805  & 0.553 &  0.087 & 1.445 & 0.014 & 1.459(87)  \\
   $^{66}$As & 0.791  & 0.570 &  0.095 & 1.456 & 0.014 & 1.470(95)  \\
   $^{70}$Br & 0.776  & 0.591 &  0.105 & 1.473 & 0.013 & 1.49(11)  \\
   $^{74}$Rb & 0.761  & 0.609 &  0.115 & 1.485 & 0.013 & 1.50(12)  \\[2mm]
\end{tabular}
\end{ruledtabular}
\end{center}
\end{table}

\item {\it Long distances, $0 \leq Q^2 \leq (0.823~{\rm GeV})^2$}: 
Integrating the long-distance amplitudes up to $Q^2 = (0.823~{\rm GeV})^2$,
where the integrand matches smoothly to the interpolation function,
they obtained a smaller value for $C_{\rm Born}^{\rm free}$,
\be
C_{\rm Born}^{\rm free} = 0.829 ,
\label{CBorn4}
\ee
than given in Eq.\,(\ref{CBorn3}).  However this smaller value is caused entirely by the
reduction in the effective upper limit to the loop integration, and is almost
completely compensated for by the consequently higher values obtained for the graph in
the other energy regimes.

\ei

In the end, Marciano and Sirlin \cite{MS06} find that the net effect of this re-evaluation
of the $\gamma W$-box axial graph is a very small reduction in the radiative
correction of $1.4 \times 10^{-4}$.  More important than this reduction, the
new method provides a more systematic estimate of the hadronic
uncertainties.  Allowing for a $\pm 10 \%$ uncertainty for the
$C_{\rm Born}^{\rm free}$ correction in Eq.\,(\ref{CBorn4}), a
$\pm 100 \%$ uncertainty for the interpolator contribution in the
intermediate region, and
$\pm 0.0001$ uncertainty from neglected higher order effects,
Marciano and Sirlin \cite{MS06} find the total uncertainty
in the radiative correction is $\pm 0.00038$.  This corresponds
to more than a factor of two reduction in the loop uncertainty
for hadronic effects ({\it cf.} Eq.\.(\ref{DRVvalu1})). 

\begin{table} [b]
\begin{center}
\caption{Calculated nuclear-structure-dependent radiative
correction,
$\delta_{NS}$, in percent units, and the component
contributions.
\label{t:tab2} }
\vskip 1mm
\begin{ruledtabular}
\begin{tabular}{lrrrr}
& & & &  \\[-2mm]
& \multicolumn{1}{c}{2002} &
\multicolumn{3}{c}{This work} \\
\cline{3-5} \\[-3mm]    
\multicolumn{1}{c}{Parent} &
\multicolumn{1}{c}{$\delta_{NS}(\% )$} & & &
\multicolumn{1}{c}{$\delta_{NS}(\% )$} \\
\multicolumn{1}{c}{nucleus} &
\multicolumn{1}{c}{Ref.\,\protect\cite{TH02}} &
\multicolumn{1}{c}{$C_{NS}^{\rm quenched}$} &
\multicolumn{1}{c}{$(q-1)C_{\rm Born}^{\rm free}$} &
\multicolumn{1}{c}{adopted} \\[1mm]
\hline 
& & & &  \\[-2mm]
\multicolumn{2}{l}{~~~~$T_z = -1$:} & & & \\
   $^{10}$C  & $-$0.360(35) & $-$1.318 & $-$0.176 & $-$0.345(35) \\
   $^{14}$O  & $-$0.250(50) & $-$0.844 & $-$0.208 & $-$0.245(50) \\
   $^{18}$Ne & $-$0.290(35) & $-$1.051 & $-$0.198 & $-$0.290(35) \\
   $^{22}$Mg & $-$0.240(20) & $-$0.750 & $-$0.213 & $-$0.225(20) \\
   $^{26}$Si & $-$0.230(20) & $-$0.705 & $-$0.227 & $-$0.215(20) \\
   $^{30}$S  & $-$0.190(15) & $-$0.557 & $-$0.242 & $-$0.185(15) \\
   $^{34}$Ar & $-$0.185(15) & $-$0.520 & $-$0.257 & $-$0.180(15) \\
   $^{38}$Ca & $-$0.180(15) & $-$0.475 & $-$0.271 & $-$0.175(15) \\
   $^{42}$Ti & $-$0.240(20) & $-$0.765 & $-$0.241 & $-$0.235(20) \\[2mm] 
\multicolumn{2}{l}{~~~~$T_z = 0$:} & & & \\
   $^{26}$Al$^m$ & 0.009(20) & 0.242 & $-$0.227 & 0.005(20) \\
   $^{34}$Cl & $-$0.085(15) & $-$0.118 & $-$0.257 & $-$0.085(15) \\
   $^{38}$K$^m$ & $-$0.100(15) & $-$0.158 & $-$0.271 & $-$0.100(15) \\
   $^{42}$Sc & 0.030(20) & 0.391 & $-$0.241 & 0.035(20) \\
   $^{46}$V  & $-$0.040(7) & 0.093 & $-$0.248 & $-$0.035(10) \\
   $^{50}$Mn & $-$0.042(7) & 0.084 & $-$0.254 & $-$0.040(10) \\
   $^{54}$Co & $-$0.029(7) & 0.112 & $-$0.261 & $-$0.035(10) \\
   $^{62}$Ga & $-$0.040(20) & 0.087 & $-$0.272 & $-$0.045(20) \\
   $^{66}$As & $-$0.050(20) & 0.010 & $-$0.278 & $-$0.060(20) \\
   $^{70}$Br & $-$0.060(20) & $-$0.085 & $-$0.283 & $-$0.085(25) \\
   $^{74}$Rb & $-$0.065(20) & $-$0.026 & $-$0.288 & $-$0.075(30) \\[2mm]
\end{tabular}
\end{ruledtabular}
\end{center}
\end{table}

\subsection{New values for $\delta_R^{\prime}$, $\DRV$ and $\delta_{NS}$}
\label{ss:col}

We maintain the traditional separation of the radiative correction into
a nucleus-dependent outer correction and a nucleus-independent inner
correction -- see Eqs.\,(\ref{dr3}) and (\ref{DR4}).  This means that the
outer correction, $\delta_R^{\prime}$, is slightly redefined and is now
written as 
\be
\delta_R^{\prime} = \frac{\alpha}{2 \pi} \left [ \overline{g}(E_m)
+ \delta_2 + \delta_3 +\delta_{\alpha^2} \right ]
\label{dr4}
\ee
where the new term, $\delta_{\alpha^2}$, simply represents the difference
between the definition of $\delta_R^{\prime}$ given in Eq.\,(\ref{dr3}) and that
given in Eq.\,(\ref{dr1}).  It is the leading-log extrapolation of the
logarithm $\ln ( m_p / 2 E_m)$, which is contained in the function
$\overline{g}(E_m)$.  Values of $\delta_{\alpha^2}$ and the redefined
$\delta_R^{\prime}$ are given in Table~\ref{t:tab1} for all the superallowed
transitions of interest.

\begin{table} [t]
\begin{center}
\caption{Corrected $\F t$ values for the thirteen best known superallowed decays,
obtained with the new correction terms presented in this work.  The experimental
$ft$ values were taken from results in our 2005 survey \cite{HT05} updated with more recent
published data \cite{Sa05, To05, Hy05, Er06a, Er06b, Bo06, Ba06, Ia06, Hy06, Bu06}.
The average $\overline{\F t}$ value and the normalized $\chi^2$ of the fit to a constant appears
at the bottom.
\label{t:Ft} }
\vskip 1mm
\begin{ruledtabular}
\begin{tabular}{lccccc}
& & & & &  \\[-2mm]
\multicolumn{2}{l}{Parent} & & & &  \\
\multicolumn{2}{l}{nucleus~~~~$ft$(s)}
& $\delta_R^{\prime}$(\%) 
& $\delta_{NS}$(\%)
& $\delta_C$(\%) 
& $\F t$(s) \\[1mm]
\hline 
& & & & &   \\[-2mm]
\multicolumn{2}{l}{~~~~$T_z = -1$:} & & & &  \\
   $^{10}$C  & 3039.5(47)  & 1.679(4) &  -0.345(35) & 0.175(18) & 3074.5(49)  \\ 
   $^{14}$O  & 3042.5(27)  & 1.543(8) &  -0.245(50) & 0.330(25) & 3071.6(33)  \\   
   $^{22}$Mg & 3052.2(72)  & 1.466(17) &  -0.225(20) & 0.380(22) & 3078.3(74)  \\ 
   $^{34}$Ar & 3052.5(82)  & 1.412(35) &  -0.180(15) & 0.665(56) & 3069.4(85)  \\ [2mm] 
\multicolumn{2}{l}{~~~~$T_z = 0$:} & & & & \\
   $^{26}$Al$^m$ & 3037.0(11) & 1.478(20) &  0.005(20) & 0.310(18) & 3072.5(15)  \\ 
   $^{34}$Cl & 3050.0(11)  & 1.443(32) &  -0.085(15) & 0.650(46) & 3071.3(21)  \\ 
   $^{38}$K$^m$ & 3051.1(10)  & 1.440(39) &  -0.100(15) & 0.655(59) & 3071.7(24)  \\ 
   $^{42}$Sc & 3046.4(14)  & 1.453(47) &  0.035(20) & 0.665(56) & 3071.2(27)  \\ 
   $^{46}$V  & 3049.6(16)  & 1.445(54) &  -0.035(10) & 0.620(63) & 3073.4(30)  \\ 
   $^{50}$Mn & 3044.4(12)  & 1.445(62) &  -0.040(10) & 0.655(54) & 3066.9(28)  \\ 
   $^{54}$Co & 3047.6(15)  & 1.443(71) &  -0.035(10) & 0.770(67) & 3066.7(33)  \\ 
   $^{62}$Ga & 3075.5(14)  & 1.459(87) &  -0.045(20) & 1.48(21) & 3073.0(72)  \\
   $^{74}$Rb & 3084.3(80)  & 1.50(12) &  -0.075(30) & 1.63(31) & 3077(13)  \\[1mm]
\cline{4-6}
& & & & &  \\[-2mm]
 & & & \multicolumn{2}{c}{Average $\overline{\F t}$} & 3071.4(8)   \\
 & & & \multicolumn{2}{c}{$\chi^2/\nu$} & 0.6   \\[1mm]
\end{tabular}
\end{ruledtabular}
\end{center}
\end{table}

The new inner correction is defined by Eq.\,(\ref{DR4}), with $C_{\rm Born}^{\rm free}$
taken from Eq.\,(\ref{CBorn4}).  With its uncertainty obtained from Marciano and
Sirlin \cite{MS06}, the result is
\be
\DRV = (2.361 \pm 0.038) \% .
\label{DRVvalu3}
\ee

It is important to note that with the re-evaluation of $C_{\rm Born}^{\rm free}$,
there is a consequent change in the nuclear-structure dependent correction
$\delta_{NS}$ given in Eq.\,(\ref{dns}).  Fortunately, the change is very
small, being
\bea
(q-1) (C_{\rm Born}^{\rm new} - C_{\rm Born}^{\rm old} )
(\frac{\alpha}{\pi} )& \simeq & -0.3 ( -0.062 ) 2.3 \times 10^{-3}
\nonumber \\
 & \simeq & 0.004 \%.
\label{changedNS}
\eea
In addition to making this change, we have also taken the
opportunity to re-evaluate $C_{NS}$ using the more recently available shell-model
effective interactions described in Sect.~\ref{sss:smc}.  Our revised $\delta_{NS}$
values are listed in Table~\ref{t:tab2}.  As in Tables\,\ref{t:dc2} and \ref{t:dc1new},
we give (in columns 3 and 4) the results from {\it one sample calculation} for each
nucleus.  Then in column 5 we present adopted $\delta_{NS}$ values that result from
our assesment of {\it all} calculations made for each decay, not just the ones shown
in columns 3 and 4; the uncertainties were chosen to encompass the spread in the
results from those calculations.  For comparison, in column 2 we list the values we
adopted for $\delta_{NS}$ in 2002 \cite{TH02}.  In all cases the new values agree with
the old ones within the quoted uncertainties.

\section{$\F t$ values, $V_{ud}$ and CKM unitarity }
\label{s:Ft}

We have calculated improved results for the correction terms $\delta_{C1}$ (see
Table\,\ref{t:dc1new}), $\delta_{C2}$ (Table\,\ref{t:dc2}) and $\delta_{NS}$
(Table\,\ref{t:tab2}); and, based on the work of Marciano and Sirlin, we have
presented revised values for $\delta_R^{\prime}$ (Table\,\ref{t:tab1}) and $\DRV$
(Eq.\,(\ref{DRVvalu3})).  We are now in a position to extract corrected $\F t$ values
from the current world data for superallowed $0^{+} \rightarrow 0^{+}$ transitions. 

\begin{figure}[b]
\epsfig{file=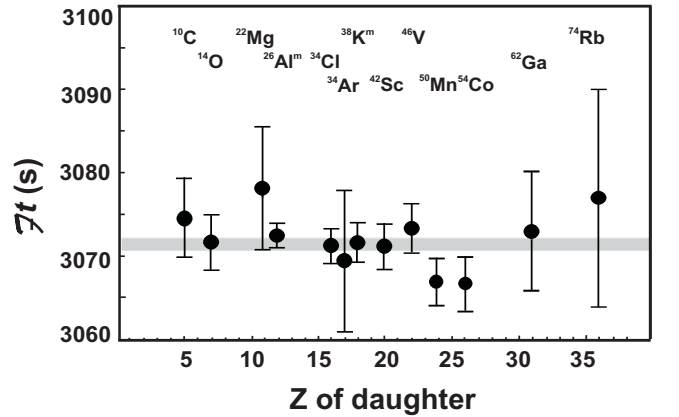,width=8.5cm}
\caption{Results for the new corrected $\F t$ values (from Table\,\ref{t:Ft}) for
the thirteen best known superallowed decays.  The corresponding uncorrected $ft$
values appear in the left panel of Fig.\,\ref{fig1}.  The shaded horizontal
band gives one standard deviation around the average $\overline{\F t}$ value.}
\label{fig2}
\end{figure}

We use the same data set as that described in Sect.\,\ref{s:sbd}: it represents an
interim update of our 2005 complete survey \cite{HT05} and includes ten additional
published measurements \cite{Sa05, To05, Hy05, Er06a, Er06b, Bo06, Ba06, Ia06, Hy06,
Bu06}.  Results are given in Table\,\ref{t:Ft} for the thirteen superallowed
transitions whose $ft$ values are known to a precision of 0.3\% or better.  The
$\F t$ values given in column 6 were obtained from the data in the preceeding columns
through the application of Eq.\,(\ref{Ftfactor}).  The corrected $\F t$ values are
also plotted in Figure \ref{fig2}.

It is clear from the normalized $\chi^2$ given on the bottom line of the table
that the statistical agreement among the $\F t$ values remains excellent.  Furthermore,
it is evident from the figure that $^{46}$V no longer shows any deviation from the
overall average as it did in Fig.\,\ref{fig1}.  However, it is equally evident
that instead the $^{50}$Mn and $^{54}$Co $\F t$ values are now low, and by amounts
that are no less statistically significant than the amount by which the $^{46}$V
value was previously high.  

Rather than being a negative result, however, this possible discrepancy offers us the
opportunity to use the cases of $^{50}$Mn and $^{54}$Co as a valuable
test of our improved calculations.  The $Q_{EC}$ value for each of them has been
measured only twice with (claimed) high precision \cite{Vo77, Ko87}, and one of these
references \cite{Vo77} also included a measurement of the $Q_{EC}$ value for
$^{46}$V, which Penning-trap measurements have recently shown \cite{Sa05,Er06a}
to be low by 2 keV -- more than three times its originally quoted
standard deviation.  If, as seems likely, the problem with the $^{46}$V measurement in
Ref.\,\cite{Vo77} is not limited to that measurement alone, then doubt is certainly cast
on the $^{50}$Mn and $^{54}$Co $Q_{EC}$-value results quoted in that reference as well.  
New Penning-trap measurements of both $Q_{EC}$ values are currently in progress \cite{Er07},
and the question should be settled shortly.  If the $Q_{EC}$ values in Ref.\,\cite{Vo77}
prove to have been too low again, then the new Penning-trap measurements will serve to
increase the $\F t$ values for $^{50}$Mn and $^{54}$Co and could well bring them into
close agreement with the average $\overline{\F t}$ value.  If so, this would add strong support to
our new calculations. 

The average corrected $\overline{\F t}$ value obtained from our new analysis, 3071.4(8)~s, is lower
by more than one standard deviation, compared to the comparable result obtained in our 2005
survey, 3072.7(8)~s.  If the new measurements do prove to increase the $Q_{EC}$ values for
$^{50}$Mn and $^{54}$Co, then this discrepancy will decrease slightly, but there is no
avoiding the fact that the inclusion of some core orbitals in the nuclear-structure-dependent
correction terms has increased the correction in a number of cases, which in turn leads to
a reduction in their $\F t$ values.  A significant change in the nuclear model has led to
a significant change -- but not a revolutionary one -- in the average $\overline{\F t}$ value.  

The new average $\overline{\F t}$ value yields a new value for $V_{ud}$ via the equation
\be
V_{ud}^2 = \frac{K}{2 \GF^2 (1 + \DRV ) \overline{\F t}},
\label{Vudeq}
\ee
where $\GF$ is the well known weak-interaction constant for the purely leptonic muon decay
\cite{PDG}.  It has been our practice when using the $\overline{\F t}$ value in this context
to add 0.85(85)\,s to its value to account for possible systematic errors in the treatment of
the radial wave function in the calculation of $\delta_C$. (This point is discussed in
detail in section III~C of Ref.\,\cite{HT05}.) Continuing this practice, we obtain the
following result for the up-down element of the CKM matrix:
\be
|V_{ud}| = 0.97418(26).
\label{Vudvalu}
\ee
This result can be compared with the value 0.97380(40), which was obtained in 2005 \cite{HT05}.
The new value is (just) within the uncertainty of the previous value, and carries an
uncertainty that is one third smaller.

The final step is to combine this new value of $|V_{ud}|$ with the other top-row elements of
the CKM matrix, $|V_{us}|$ and $|V_{ub}|$, to test the unitarity of the matrix.  Taking the
values of the latter two elements from the 2006 Particle Data Group review \cite{PDG} we
obtain the stunning result
\be
|V_{ud}|^2 + |V_{us}|^2 + |V_{ub}|^2 = 1.0000 \pm 0.0011.
\label{unitarity}
\ee
Unitarity is fully satisfied with a precision of 0.1\%.
\\[3mm]

\section{Conclusions}
\label{s:Conc}

We have presented new calculations of the nuclear-structure-dependent corrections
to superallowed $0^{+} \rightarrow 0^{+}$ nuclear $\beta$ decay.  The calculations
incorporate core orbitals in the shell model in cases where independent experimental
information indicates that they are required.  Where possible, they also make use of
effective interactions that have been published since our previous calculation of
these correction terms \cite{TH02}.  As in that work, we have included twenty
transitions in our calculations, thirteen that are by now rather well measured and
seven more that are likely to be accessible to precise measurements in the future.

The agreement among the corrected $\F t$ values for the thirteen well measured cases
is very good, although there is a possible small discrepancy for the cases of $^{50}$Mn
and $^{54}$Co.  A new Penning-trap measurement of the $Q_{EC}$ values for these two
transitions is expected in the near future, and its effect on this discrepancy could
serve to test the validity of our calculations.

With our new corrections, the value of $|V_{ud}|$ is increased by 0.04\%, or by one
standard deviation of the previous result \cite{HT05}.  With the new value, the sum of
squares of the top-row elements of the CKM matrix is in perfect agreement with
unitarity.

The improved calculations presented in this work were inspired by the remarkable recent
improvements in experimental precision, particularly in the measurement of the $^{46}$V
$Q_{EC}$ value.  The only way that the calculated corrections can be
tested and improved is by such precise measurements, both on the currently well-known
transitions and on other as-yet-unstudied superallowed transitions that have larger
calculated corrections.  If the calculated correction terms replace the significant
scatter in the measured $ft$ values (see the left panel in Fig.\,\ref{fig1}) with a
set of self-consistent corrected $\F t$ values, then they can surely be relied upon
to produce a secure value for $|V_{ud}|$.  The present calculations testify to the
value of increased experimental precision. 
 
\acknowledgments

The work of JCH was supported by the U. S. Dept. of Energy under Grant
DE-FG03-93ER40773 and by the Robert A. Welch Foundation under Grant A-1397.
IST would like to thank the Cyclotron Institute of Texas A \& M University
for its hospitality during annual two-month summer visits.

\end{document}